\documentclass{amsart}

\usepackage{amssymb}
\usepackage{amsmath}
\usepackage{amsthm}
\usepackage{stmaryrd}
\usepackage{graphicx}
\usepackage{epsfig}

\DeclareMathOperator{\var}{Var}

\def\defby{=:}
\def\bydef{:=}
\def\build#1_#2^#3{\mathrel{\mathop{\kern 0pt#1}\limits_{#2}^{#3}}}
\def\tend#1#2#3{\build\hbox to 12mm{\rightarrowfill}_{#1\rightarrow
#2}^{#3}}

\def\reff#1{(\ref{#1})}

\newcommand{\one}{\mbox{\tiny1}}
\newcommand{\po}{\mbox{\tiny+1}}
\newcommand{\mo}{\mbox{\scriptsize-\tiny1}}
\newcommand{\ze}{\mbox{\tiny0}}

\newcommand{\mtw}{\mbox{\scriptsize-\tiny2}}

\newcommand{\Ta}[2]{T^{(a)}_{#1\rightarrow #2}}
\newcommand{\x}[2]{X^{(a)}_{#1}({#2})}
\newcommand{\xa}{X^{(a)}}
\newcommand{\Xa}[1]{X^{(a)}({#1})}

\newcommand{\X}[3]{X^{#1}_{#2}({#3})}
\newcommand{\T}[2]{T_{#1\rightarrow #2}}

\newcommand{\Ts}[2]{T^{#1}_{#2}}
\newcommand{\Ss}[2]{S^{#1}_{#2}}

\newcommand{\pia}[1]{\pi^{(a)}\left({#1}\right)} 
\newcommand{\E}[1]{\mathbb{E}\left[{#1}\right]}
\newcommand{\pr}[1]{\mathbb{P}\left({#1}\right)}
\newcommand{\p}[2]{p\left( {#1}, {#2} \right)}

\newcommand{\linf}[1]{\xrightarrow[#1\to\infty]{}}

\newcommand{\Bx}{\llbracket x,a\rrbracket}
\newcommand{\LBx}{\llbracket b,x\rrbracket}
\newcommand{\LB}[2]{\llbracket #1,#2\rrbracket}
\newcommand{\EE}[1]{\mathbb{E}\bigl[{#1}\bigr]}
\newcommand{\EEE}[1]{\mathbb{E}\Bigl[{#1}\Bigr]}
\newcommand{\prr}[1]{\mathbb{P}\Bigl({#1}\Bigr)}
\newcommand{\alf}{\mathcal{A}_a}
\newcommand{\alff}{\mathcal{A}}
\newcommand{\high}{\mathcal{H}_a}
\newcommand{\med}{\mathcal{K}}
\newcommand{\low}{\ell}
\newcommand{\scale}{\Delta_a}
\newcommand{\qa}{Q^{(a)}}

\newcommand{\Tal}[2]{\check T^{(a)}_{#1\rightarrow #2}}
\newcommand{\Tar}[2]{\hat T^{(a)}_{#1\rightarrow #2}}
\newcommand{\xal}{\check X^{(a)}}
\newcommand{\xar}{\hat X^{(a)}}
\newcommand{\ppl}{\check p}
\newcommand{\qql}{\check q}
\newcommand{\rrl}{\check r}
\newcommand{\ppr}{\hat p}
\newcommand{\qqr}{\hat q}
\newcommand{\rrr}{\hat r}
\newcommand{\pial}[1]{\check\pi^{(a)}\left({#1}\right)} 
\newcommand{\piar}[1]{\hat\pi^{(a)}\left({#1}\right)} 

\newcommand{\uuu}{\mathcal{U}_{x\to y}}
\newcommand{\vvv}{\mathcal{V}^k_{x\to y}}
\newcommand{\rev}{\mathcal{R}}
\newcommand{\TT}{\widetilde T_{x\to y}}
\newcommand{\TTr}{\widetilde T_{y\to x}}

\newtheorem{theorem}{Theorem}[section]
\newtheorem{proposition}[theorem]{Proposition}
\newtheorem{lemma}[theorem]{Lemma}
\newtheorem{corollary}[theorem]{Corollary}
\newtheorem{definition}[theorem]{Definition}
\newtheorem{remark}[theorem]{Remark}

\begin{document}

\title{Abrupt convergence and escape behavior\\ for birth and death chains}
\author{J. Barrera}
\address{
School of Engineering, Universidad Adolfo Iba\~nez\\
Avda. Diagonal Las Torres 2640, Pe\~nalolen, Santiago,  Chile.}
\email{javiera.barrera@uai.cl}

\author{O. Bertoncini}
\address{
Laboratoire de Math\'ematiques Rapha\"el Salem\\
UMR 6085 CNRS-Universit\'e de Rouen\\
 Avenue de l'Universit\'e, BP.12, Technop\^ole du Madrillet, F76801 Saint-\'E‰tienne-du-Rouvray.}
\email{ olivier.bertoncini@univ-rouen.fr}

\author{R. Fern\'andez}
\address{
Laboratoire de Math\'ematiques Rapha\"el Salem\\
UMR 6085 CNRS-Universit\'e de Rouen\\
 Avenue de l'Universit\'e, BP.12, Technop\^ole du Madrillet, F76801 Saint-\'E‰tienne-du-Rouvray.}
\email{ roberto.fernandez@univ-rouen.fr}

\date{\today}

\maketitle

\begin{abstract}
We link two phenomena concerning the asymptotical behavior of sto-chastic processes: (i) abrupt convergence or cut-off phenomenon, and (ii) the escape behavior usually associated to exit from metastability.  The former is characterized by convergence at asymptotically deterministic times, while the convergence times for the latter are exponentially distributed.  We compare and study both phenomena for discrete-time birth-and-death chains on $\mathbb{Z}$ with drift towards zero.  In particular, this includes energy-driven evolutions with energy functions in the form of a single well.    Under suitable drift hypotheses, we show that there is both
an abrupt convergence towards zero and escape behavior in the other direction.  Furthermore, as the evolutions are reversible, the law of the final escape trajectory coincides with the time reverse of the law of cut-off paths.  Thus, for evolutions defined by one-dimensional energy wells with sufficiently steep walls,  cut-off and escape behavior are related by time inversion.

{\bf keywords}: cut-off \and metastability \and hitting time \and exit-times \and reversibility  
\end{abstract}


\section{Introduction}
\begin{figure}[htbp]
\centering
\scalebox{0.30}{\input{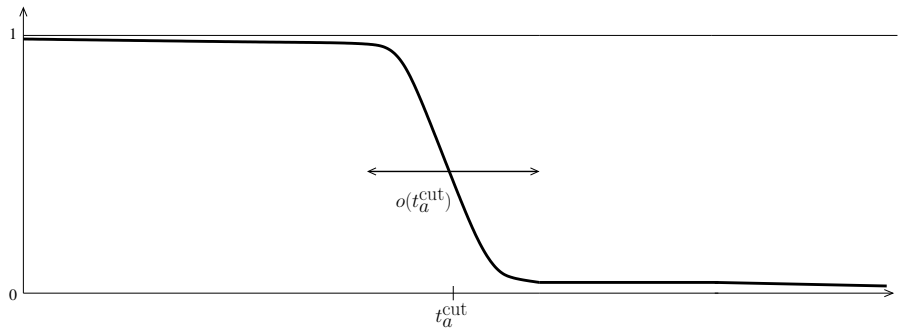}}
\caption{Distance to equilibrium during abrupt convergence}
\label{fig:1}
\end{figure}

The \emph{cut-off} phenomenon, first identified and formalized in the early eighties~\cite{DiaAlShCST,DiaAlSUT}, is by now a well studied feature of Markov processes (see \cite{DiacGrpRepr,Diacut,SClecfmc,SCrwfg} for reviews).  The phenomenon refers to an asymptotically drastic convergence of a family of stochastic processes $X^{(a)}$ labeled by some parameter $a$.  As $a\to\infty$, a suitable distance between the laws $\pr{X^{(a)}(t) \in \bullet}$ and the corresponding invariant measures $\pia{\bullet}$ converges, in macroscopic time units, to a step function centered at \emph{deterministic} times $t^{\rm cut}_a$.  More precisely, the function $a\to t^{\rm cut}_a$ is such that the distance is asymptotically maximal for times smaller than $t^{\rm cut}_a-o(t^{\rm cut}_a)$ and asymptotically zero for times larger than $t^{\rm cut}_a+o(t^{\rm cut}_a)$ (Figure \ref{fig:1}). 
The term \emph{cut-off} is naturally associated to such an ``all/nothing'' or ``$1/0$'' behavior, but it has the drawback of being used with other meanings in statistical mechanics and theoretical physics.   Alternative names have been proposed, including \emph{threshold phenomenon}~\cite{DiaAlSUT} and \emph{abrupt convergence}~\cite{YcAlgomkv}.  It has also been called \emph{abrupt switch} in a precursor work~\cite{AldousRW} based on mixing times rather than distances between measures.

Despite growing interest among probabilists, cut-off convergence has been largely ignored within mathematical statistical mechanics (\cite{levlucper07} is the only exception we know of, besides our previous announcement~\cite{berbarfer08} and the thesis of one of us~\cite{olivphd}).  As a tentative justification of this situation, we stress the fact that cut-off studies have often been performed in a relatively elaborated setting that involves a certain amount of technical complexity.  Indeed, cut-off was initially studied in evolutions defined by group actions (eg.\ card shuffling)~\cite{DiaAlShCST,DiaAlSUT}. Emphasis was, therefore, put into Fourier techniques~\cite{DiacGrpRepr}.  Subsequently~\cite{DiaAlSUT,DiacTrail,DiacRWaris,DiacTopto,DiacAsymp,DiacTimeto},
the phenomenon was associated to the existence of ``strong uniform (coupling) times'' and to symmetry and multiplicity properties of second eigenvalues of transition matrices, see \cite{Diacut} for a review. It has also been associated to exponential speed of convergence in  \cite{YcCutSMC} ,  \cite{BaLaYc} and \cite{LaOU}.  
 This fostered the use of notions and techniques issued from diverse mathematical fields, ranging from  combinatorics to functional analysis, see~\cite{SClecfmc,SCrwfg}.  The mathematical richness of these studies may have prevented potential audiences from fully appreciating the probabilistic meaning of the phenomenon.  

In all these references, the existence of the phenomenon is proven through a precise estimation of the sequence $ t^{\rm cut}_a$ of cut-off times.  This precision comes at a high technical price and is largely responsible for the variety of treatments found in the literature. It is clear that the theory would benefit from the disentanglement of the proof of  the \emph{existence} of a cut-off from the more delicate determination of the actual cut-off times. A contribution in this direction is the work done in relation with the conjecture formulated by Y. Peres in 2004.
The conjecture states that in different natural settings a necessary  and sufficient condition for a process to have cut-off is that the product of mixing time and the spectral-gap  tends to infinity. It was proved that the conjecture is true for Markov chains for the $L^p$ norm with $p>1$ in \cite{ChenSaloffCoste}. It was also proved to be true for birth and death chains in separation distance in \cite{DiaconisSaloffCoste} and in total variation distance in \cite{PeresB-D}.  These articles have been a big contribution to the understanding of the phenomenon, yet the condition to observe the phenomenon still relies in the spectral structure of the chain and the behavior of the distance. In our opinion, this work should be complemented by an approach based on the analysis of
typical trajectories, which would reveal, more clearly, the underlying probabilistic
mechanism leading to cut-off among all possible patterns of convergence. The present work
points in this direction.
$~$\\

 An alternative, more probabilistic approach proposed in the last decade~\cite{YcCutSMC,YcartMCMC,Ycutexple,YcartMart} offers some advantages regarding the above criticism.  In this approach, cut-off is associated to the existence of an appropriate drift in stochastic processes or sampling schemes, and arguments  are based on the behavior of hitting times~\cite{LaOU,LaYc,BaLaYc}.  Furthermore, the approach includes some general criteria for the existence of cut-off as illustrated by  Proposition \ref{condsuffcut} below (taken from~\cite{YcartMart}).  
The resulting description offered some tantalizing contrast with another, quite different type of convergence associated to metastability.  This observation triggered our research.

\begin{figure}[htbp]
\centering
\scalebox{0.40}{\input{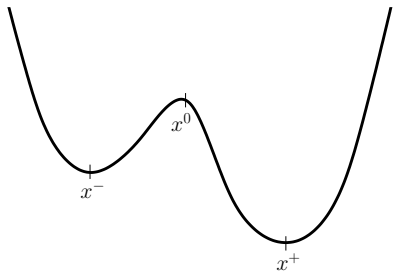}}
\caption{The local minimum at $x^-$ leads to metastability}\label{fig:2}
\end{figure}

From a physical point of view, a system exhibits metastability if it stays trapped for exceptionally long times in a state different from true equilibrium (eg.\ supercooled water).  The final transition to equilibrium is rather sudden and happens at ``unpredictable'' times.  Mathematically, it is the behavior expected for evolutions designed to find a minima $x^+$ of some energy function $H(x)$, when the system is ``captured'' by a local minimum at $x^-$ (Figure~\ref{fig:2}).

Unlike cut-off, metastable behavior has been thoroughly studied within rigorous statistical mechanics.
The basic scenario was described in a seminal paper~\cite{CGOV} that is contemporary of the first cut-off papers:  A system driven downhill towards $x^-$ will stay for a long time around this point until suddenly, at some random instant $t^{\rm esc}$, an escape trajectory unfolds, that takes it over the barrier at $x^0$ and down to the true minimum $x^+$.  The law of $t^{\rm esc}$ converges  to an exponential as the  ``barrier height'' $x^0-x^-$ grows in a suitable manner.  In a typical trajectory, the distance to $x^+$ follows curves similar to those of Figure~\ref{fig:1}, except that the jump-times $t_a$ are exponentially distributed rather than determinist.  The parameter $a$ diverges with the height and/or steepness of the small well.  

This picture, borrowed from the theory of randomly perturbed dynamical systems~\cite{FW}, has been called ``path approach to metastability''.  It has subsequently been refined and systematized~\cite{OS1,OS2,ScoppMMCren};  
see~\cite{OStypexpath,OStunntime,OliVares} for overviews.  
Escape from metastability happens through a very narrow choice of trajectories, known as the \emph{exit tube}, which has been determined in a very precise manner for a number of spin dynamics~\cite{Schapproach,Schpatesc,NevSchcritdrop,NevSchbehavdrop,KOdropdyn,dHOS}.  
A more recent and efficient alternative rigorous treatment of metastability is based on potential theory~\cite{bovetal00,bovetal01,bovetal01b,bovfran02,BovGlauDyn,BovMetaRDP1,BovMetaRDP2}.  The approach relies on an optimization principle which allows the determination of metastable behavior through suitable trial functions, without an in-depth study of trajectories (see~\cite{BovMetageing,bov05,BovMetaPTA} for reviews).  

Cut-off and escape behavior show, therefore, both similar and contrasting features.  On the one hand, 
both are \emph{asymptotic} phenomena: they show up when a suitable parameter is sufficiently large and their very definitions rely on limit properties.  On the other hand, while both convergences are sudden, the laws of the convergence times are antipodal: deterministic for cut-off and exponential (randomness without memory) for escapes.  There are also differences in focus:  cut-off studies emphasize distance between measures, while exit times are the object of metastability studies.  The message of the present paper, however, is that in some natural sense they are often intertwined phenomena.  

In order to exhibit this intertwining, we adopt a relatively simple, but sufficiently rich, framework where both the probabilistic approach to cut-off of~\cite{YcCutSMC,YcartMCMC,Ycutexple,YcartMart} and the path approach to metastability of~\cite{FW} can be applied.  We consider birth-and-death chains with a drift (towards the origin) and focus on hitting times for excursions along and against the drift.  
Under reasonable hypothesis on the drift, we show that cut-off behavior characterizes the first type of excursions, where the latter requires exponential escape times.  If an energy function is associated to the drift, this results can be interpreted as saying that, if the well is sufficiently steep or high (and its walls are sufficiently smooth), climbing along a wall requires an exponential time while descents are abrupt.   Furthermore, by reversibility, both types of excursions are related by time inversion:  the law of  escape trajectories equals the time reverse of the law of cut-off paths. In particular, our results apply to rather general recurrent random walks and to the Ehrenfest model.  The latter is a kind of border line case, as it is associated to a well whose walls are not as steep as required in standard metastability studies.

Our results explain, in part, why cut-off phenomena has been largely ignored by physics-minded researchers: it corresponds to the ``trivial" downhill evolution of a system falling in an energy well.   At the same time, our work suggests a more detailed decomposition of metastable evolutions.  Coming back to Figure~\ref{fig:2}, the escape from the local minimum at $x^-$ to the absolute one at $x^+$ can be decomposed in three stage: (i) the exit from $x^-$ to the left-vicinity of $x^0$, (ii) the transition to the right of $x^0$, and (iii) the descent to $x^+$.  The exponential time typical of the transitions to equilibrium ---from $x^-$ to $x^+$--- is, in fact, determined by stage (i).  The time scale of the other two stages is so much smaller that it dissapears under the usual time rescaling.  Evolutions in more complicated energy landscapes can, in principle, be decomposed into a sequence of exponentially distributed uphill pieces followed by cut-off-like downhill excursions.  Only the uphill stages survive the usual time rescaling, and the resulting theory should merge with existing metastability studies~\cite{OS1,OS2,ScoppMMCren}.

Furthermore, the equality, modulo time-reversal, of the laws of up- and downhill trajectories clarifies the escape scenario:  The final excursion taking the system from the bottom to the top of a well is the reversed of the cut-off trajectory that would take it in the opposite direction.  Hence, this upwards motion is done in one stroke, almost without midway hesitations, in an asymptotically deterministic manner.  The exponential law of the exit time is due to the fact that the system spends an exponential time around the bottom of the well till the atypical uphill excursion takes place.   

Our results suggest a profitable association between the cut-off and metastability research communities.
We warn, however, that the intertwining between cut-off and exponential escape times is by no means universal.   Exponential escape times are a more robust phenomenon that requires uphill time scales much larger than downhill ones.  But the latter need not be cut-off like.  Wells whose walls are rugged, or not strictly monotone, yield examples where the downhill time is not abrupt while the uphill law remains exponential.  In this sense, knowing that downhill excursions exhibit cut-off conveys more detailed information than knowing that the opposite evolution is escape-like.   Nevertheless, it is reasonable to expect that methods devised to study one of the phenomena might be adaptable to the study of the other.  This can be of particular interest, for instance to study evolutions involving bidimensional ---or multidimensional--- wells.


\section{Definitions and results}\label{Def&Results}
\subsection{The two types of behavior.}  Cut-off and escape behavior, will be studied at the level of hitting times.  Both types of behavior are asymptotic, in the sense that they are characterized by what happens when a certain parameter $a$ diverges.  Let us start with the relevant definitions.
\begin{definition}\label{def:lr0}
\begin{itemize}
\item[(i)] A family of random variables $U^{(a)}$ \emph{exhibits cut-off behavior at mean times} if 
\begin{equation}
\frac{U^{(a)}}{\E{U^{(a)}}} \;\tend{a}{\infty}{\rm Proba}\; 1\;.
\label{eq:0.1}
\end{equation}
[equivalently, $\lim_{a\to\infty}\mathbb{P}\bigl(U^{(a)} > c\, \mathbb{E}[U^{(a)}]\bigr)= 1$ for $c<1$ and $0$ for $c>1$].
\item[(ii)] A family of random variables $V^{(a)}$  \emph{exhibits escape-time behavior at its mean times} if 
\begin{equation}
\frac{V^{(a)}}{ \E{V^{(a)}}} \;\tend{a}{\infty}{\mathcal{L}}\; \exp(1) \;.
\label{eq:0.2}
\end{equation}
\end{itemize}
\end{definition}

The original definition of cut-off~ \cite{AldousRW,DiaAlShCST,DiaAlSUT} is in terms of the variational distances of invariant and process measures.  We pass to variables because the comparison with escape behavior is possible only at the level of hitting times (the measures asymptotically associated to metastable escapes are invariably reduced to delta-like measures).  The connection between both types of definition ---in terms of measure-theoretical distances and of hitting times--- has been explored in~\cite{YcartMart}, where the following is concluded:
\begin{itemize}
\item[(i)] If the variable $U^{(a)}$ correspond to an absorption time, both definitions are equivalent,
\item[(ii)] For a positive recurrent Markov chain $X_a(t)$ on $\mathbb{N}$ with invariant probability measure $\pi$ and initial state $a$, the chain admits a cut-off at time $t_a$ if and only if its associated absorbed chain with invariant measure $\delta_0$ admits a cut-off at the same time.
\end{itemize}


\subsection{The models.} Each of the models considered in the sequel corresponds to a family $\bigl\{\xa:a\in\mathbb{N}\bigr\}$ of  irreducible discrete birth-and-death chains $\xa=\bigl\{\Xa{t}: t\in \mathbb{N}\cup\{0\}\bigr\}$ on intervals $\llbracket b,a\rrbracket\bydef\{b,b-1,\ldots,a\}\subset\mathbb{Z}$.  Without loss of generality, we choose $a>0$ and $b\le 0$. The right endpoint $a$ is the diverging parameter that will reveal the asymptotic behavior.  We shall consider two cases: (i) \emph{half well}: $b=0$, which corresponds to a well in the form of a wedge with a vertical wall at the origin, and (ii) \emph{full well}: $b$ tends to minus infinity as $a$ grows, at a proportional rate (most parameters involved in the definition of the processes will be $a$-dependent but, to unclog formulas, we shall only exhibit this dependency selectively).  Chains with initial state $z$ will be denoted $\xa_z$.

Each family of chains is  characterized by transition probabilities $p_x$ to the right and $q_x$ to the left:
$0<p(x,x+1)\defby p_x$, $b\le x\le a-1$, and $0<p(x,x-1)\defby q_x$, $b+1\le x\le a$.  All other transition probabilities are zero.  The ``waiting probabilities'' $r_x\bydef 1-p_x-q_x$, $b\le x\le a$ may be non-zero, but they play little role in the sequel.   The different $p$'s and $q$'s may depend on $a$.  For each $a$ the chain is positive recurrent and admits a unique invariant probability measure $\pi^{(a)}$ which is also reversible and can be explicitly written in terms of the transition probabilities:
 \begin{equation}\label{invmes}
\pia{x}\;=\;\left\{\begin{array}{ll}
\prod_{i=\one}^{x}\, \frac{p_{i\mo}}{q_i} \,\pia{0} & x\in\LB1a\\
\prod_{i=x}^{-\one}\, \frac{q_{i\po}}{p_i} \,\pia{0} & x\in\LB{b}{-1}
\end{array}\right.
\end{equation}
This expression follows readily from the reversibility condition $\pia{x}\,q_x=\pia{x-1}\,p_{x-1}$. [Note that the values of $r_x$ appear only through the normalization constant $\pia0$, and hence they are largely irrelevant for the rest of our discussion.]

We shall focus on the behavior of hitting times
\begin{equation}
\Ta{x}{y} \;=\; \min\bigl\{ t\ge 0: \x{x}{t} =y\bigr\}\;.
\end{equation}
More specifically, we shall consider  $\Ta{a}{\ze}$, $\Ta{b}{\ze}$, $\Ta{\ze}{a}$, $\Ta{\ze}{b}$ and
$\Ta{\ze}{\{a,b\}}=\min\bigl(\Ta{\ze}{a}, \Ta{\ze}{b}\bigr)$. 
Notice that according to this definition $\Ta{x}{x}=0$ for all $x$ in $\llbracket b,a\rrbracket$.\\

\subsection{The drift condition.}  We impose the following drift condition towards $0$.  For the sake of precision, we state it exactly as needed in proofs.  The ensuing comments  relates it with the picture in terms of energy wells.

\begin{definition}
A family $\xa$ of birth-and-death process has a \emph{strong drift to the left towards 0} (we note $0^+$-SD) if 
\begin{itemize}
\item[(i)] The transition probabilities $q_x$ satisfy
\begin{equation}
\inf_{a\in\mathbb{N}} \inf_{x\in\llbracket 1, a\rrbracket} q_x \defby K_q > 0\;.
\label{eq:rl1}
\end{equation}
\item[(ii)] The constant
\begin{equation}\label{eq:rlH}
K_a\;\bydef\; \sup_{x\in\LB1a} q_x\,\EE{\Ta{x}{x\mbox{\tiny\em-1}}}
\end{equation} 
satisfies 
\begin{equation} \label{H}
 \frac{K_a^2}{\EE{\Ta{a}{\ze}}}\linf{a}0 \;
\end{equation}
\end{itemize}
Likewise, it has a \emph{Strong Drift to the right towards 0} (equivalently $0^-$-SD) if 
\begin{itemize}
\item[(i)] 
\begin{equation}
\inf_{a\in\mathbb{N}} \inf_{x\in\llbracket b, -1\rrbracket} p_x \defby K_p > 0\;.
\label{eq:rl2}
\end{equation}
\item[(ii)]
\begin{equation}\label{eq:rlHb}
K_b\;\bydef\;\sup_{x\in\LB{b}{-1}} p_x\,\EE{\Ta{x}{x\mbox{\tiny\em+1}}}
\end{equation} 
satisfies
\begin{equation} \label{Hb}
 \frac{K_b^2}{\EE{\Ta{b}{\ze}}}\linf{a}0\;. 
\end{equation}
\end{itemize}
The family $\xa$ is said to \emph{satisfy the (bilateral) Strong Drift condition towards 0 } (0-SD) if it has both $0^+$-SD  and $0^-$-SD .
\end{definition}
\medskip

\medskip
Let us discuss the meaning of the 0-SD condition.  All of the expressions given above in terms of hitting times have equivalent forms in terms of the invariant measures $\pi^{(a)}$.  The connection is provided  by the following well known expressions
\begin{equation}\label{eq:rltx}
\EE{\Ta{x}{x\mbox{\tiny\em-1}}}=\frac{\pi^{(a)}(\Bx)}{q_x\,\pi^{(a)}(x)} 
\quad;\quad\EE{\Ta{x}{x\mbox{\tiny\em+1}}}=\frac{\pi^{(a)}(\LBx)}{p_x\,\pi^{(a)}(x)} 
\end{equation}
which, for completeness, are proven in Section \ref{Htim} below.
As a consequence,
\begin{equation}
K_a \;=\; \sup_{x\in\llbracket 0, a\rrbracket} \frac{\pi^{(a)}(\Bx)}{\pi^{(a)}(x)}
\quad;\quad K_b \;=\; \sup_{x\in\llbracket b, -1\rrbracket} \frac{\pi^{(a)}(\LBx)}{\pi^{(a)}(x)} \;.
\label{eq:rl10n}
\end{equation}
These expressions show that $K_a$ and $K_b$ provide an exponential bound for the tail of $\pi^{(a)}$.  Indeed, by the leftmost identity in \reff{eq:rl10n},
\begin{equation}
\frac{\pia{\LB{x+1}{a}}}{\pia{\Bx}}\;=\; 1-\frac{\pia{x}}{\pia{\Bx}} \; \le\; 1 - \frac{1}{K_a}\;.
\end{equation}
Hence, upon iteration,
\begin{equation}\label{decexp_a}
\pi^{(a)}(x)\leq \pi^{(a)}(\Bx)\leq e^{-\alpha_a x}\quad \text{ with } \quad\alpha_a=-\log(1-1/K_a)\;,
\end{equation}
for $0\le x\le a$.  The rightmost identity in \reff{eq:rl10n} yields the same inequalities for $b\le x\le0$, with $b$ in place of $a$. 

The 0-SD condition can be transcribed  in terms of energy profiles by proposing a function $H(x)$ such that $p(x,y)\propto \exp\bigl\{-[H(x)-H(y)]\bigr\}$.  Then, $H(x)-H(x-1) = -(1/2)\log(p_{x-1}/q_x)$ and, by \reff{invmes}, $\pi^{(a)}(x)/\pi^{(a)}(0) = \exp\bigl\{\pm[H(x)-H(0)]\bigr\}$ with the upper sign for $b\le x\le0$ and the lower one for $0\le x\le a$.  Thus, from \reff{decexp_a},
\begin{equation}
H(x) - H(0) \;\ge\; \left\{\begin{array}{ll}
\alpha_a \, x & \mbox{for}\quad 0\le x\le a\\
\alpha_b \, x & \mbox{for}\quad b\le x\le 0\;.
\end{array}\right.
\label{profile}
\end{equation} 
In the standard setting for energy-driven dynamics, the steepness $\alpha$ of the walls is assumed to \emph{grow} as the parameter playing the role of $a$ (often the inverse temperature) diverges.  Here we do not require such a drastic behavior.  In fact, $\alpha$ is allowed \emph{to go to zero} (that is, $K_a\to\infty$) as long as \reff{H} and \reff{Hb} remain valid.  This generality allows our results to be applicable to the Ehrenfest model.  Condition \reff{H}  limits $K_a$ to grow sub-linearly with $a$, and \reff{Hb} implies an analogous limitation for $K_b$ as a function of $b$.  Indeed, \reff{eq:rltx} implies that 
$\EE{\Ta{a}{\ze}}\le K_q\,K_a\,a$ and, hence, the validity of \reff{H} means that
\begin{equation}
\frac{K_a}{a} \;\le\; \frac{K_q\,K_a^2}{\EE{\Ta{a}{\ze}}}\;\linf{a}\;0\;.
\label{eq:rl11}
\end{equation}

Note that, in particular, a family $\xa$ has  $0^+$-SD if $K_a$ is bounded uniformly in $a$.  This is because $\EE{\Ta{a}{\ze}}\ge a\to\infty$.  Likewise, it has $0^-$-SD if $\sup_b K_b<\infty$ as long as $b$ diverge with $a$.

\subsection{Main result and applications.} \label{ssect:rr1}
To state the main result we need one further piece of notation.  For $x,y\in\llbracket b,a\rrbracket$, let
$\tau^{(a)}_{x\to x,y} = \sup\bigl\{0<t<T^{(a)}_{x\to y} : X_x^{(a)}(t)=x\bigr\}$ be the instant of the last visit to $x$ before hitting $y$ and $\widetilde T^{(a)}_{x\to y} = T^{(a)}_{x\to y}-\tau^{(a)}_{x\to x,y}$ the time needed to hit $y$ after last visit to $x$.  The following results apply also to the particular case $b=0$ ---which we call \emph{half well}.  In this case ---which constitutes, in fact, the building block of our arguments--- every condition involving $b$ is understood to be vacuous.

\begin{theorem}\label{theo:0}
Let $\xa$ be a family of irreducible birth-and-death chains on $\LB{b}{a}$.  Then:
\begin{enumerate}
\item By reversibility, $ \widetilde T^{(a)}_{a\to \mbox{\tiny\em0}} \stackrel{\mathcal{L}}{=} \widetilde T^{(a)}_{\mbox{\tiny\em0}\to a}$ and $\widetilde T^{(a)}_{b\to \mbox{\tiny\em0}} \stackrel{\mathcal{L}}{=} \widetilde T^{(b)}_{\mbox{\tiny\em0}\to b}$. (More generally, $ \widetilde T^{(a)}_{x\to y} \stackrel{\mathcal{L}}{=} \widetilde T^{(a)}_{y\to x}$ for any $x,y\in\LB ba$.)
\item If $\xa$ has a SD to the left of 0 ($0^+$-SD), resp.\ to the right of 0 ($0^-$-SD), the random variables $\Ta{a}{\mbox{\tiny\em0}}$, resp.\ $\Ta{b}{\mbox{\tiny\em0}}$, exhibit cut-off behavior at mean times.
\item Assume $\xa$ satisfies the 0-SD condition and consider the conditions:
\begin{equation}
\label{eq:rlns1}
 \frac{\EE{\Ta{b}{\ze}}}{\EE{\Ta{\ze}a}} \;\linf a0\quad ,\quad  \frac{\EE{\Ta{a}{\ze}}}{\EE{\Ta{\ze}b}} \;\linf a0\;.
\end{equation}
\begin{enumerate}
\item If the leftmost, resp.\ the rightmost, condition is satisfied, then 
the random variables $\Ta{\mbox{\tiny\em0}}{a}$, resp.\ $\Ta{\mbox{\tiny\em0}}{b}$, exhibit escape-time behavior at mean times. 
\item If both conditions are satisfied, 
the random variables $\Ta{\mbox{\tiny\em0}}{\{a,b\}}$ exhibit escape-time behavior at mean times. 
\end{enumerate}
\end{enumerate}
\end{theorem}

Note that, for processes with strong drift towards zero, the conditions \reff{eq:rlns1}  are satisfied if there exist strictly positive constants $c_1$ and $c_2$ such that
\begin{equation}
\label{eq:rlns}
c_1\;\le\; \frac{\EE{\Ta{a}{\ze}}}{\EE{\Ta{b}{\ze}}} \; \le \; c_2\;,
\end{equation}
As commented in Section \ref{sec:full}, a weaker sufficient condition ---also based on magnitudes involving trajectories on one side of the well--- is obtained by replacing the hitting times in \reff{eq:rlns1}  by those for the corresponding ``half-well'' models (see Remark \ref{rem:rr1} below).  In particular, requirements \reff{eq:rlns1} are automatically satisfied for symmetric wells.

Let us mention a few applications.
\medskip\par\noindent
\emph{Simple (non-symmetric) random walk.} The simplest application is a random walk with transition rates that only depend on the sign of $x$ and are independent of $a$: $p_x=p^{+}$ and $q_x=q^{+}$ for $x>0$, while $p^{-}$ and $q_x=q^{-}$ for $x<0$.  Formulas \reff{eq:rltx} yield, in this case, $K_a=q^+/(q^+-p^+)$ and $K_b=p^-/(p^--q^-)$ (hence $K_q\le 2$).  Thus, the resulting family $\xa$ has a $0^+$-SD iff $q^+>p^+$.  Likewise, it has a $0^-$-SD iff \ $p^->q^-$ and $b\to\infty$ with $a$. (That is, SD = positive recurrence.) For these walks, the difference between the mean times towards and away from zero is noticeable.  Indeed, $\EE{\Ta{a}{\ze}}\sim a/(q^+-p^+)$ while, for the half well ($b=0$), $\EE{\Ta{\ze}{a}}\sim (q^+/p^+)^a$.

\medskip\par\noindent
\emph{Random walk with varying rates.} More generally, Theorem \ref{theo:0} applies to random walks with
\begin{equation}
\label{eq:lr12}
q^+\bydef \min_{x\in \LB1a} q_x > \max_{x \in\LB0a} p_x \defby p^+
\quad,\quad
p^-\bydef \min_{x\in \LB{b}{-1}} p_x > \max_{x \in\LB{b}{-1}} q_x \defby q^-
\end{equation} 
and $b$ diverging with $a$.  The values of $K_a$ and $K_b$ are as in the previous example and, as there, $\EE{\Ta{a}{\ze}}\leq {\rm const} \times a$ and $\EE{\Ta{\ze}{a}}\geq {\rm const}\times(q^+/p^+)^a$ (for $b=0$).  Furthermore, Theorem \ref{theo:0} holds even if the wall has a ``flat'' bottom. More precisely, it holds if $q_x=p_x=1/2$ for $x\in\LB{d^-_a}{d^+_a}$ for some $d^-_a\le 0$ and  $d^+_a> 0$, as long as
\begin{equation}
\label{eq:lr14}
\frac{(d^+_a)^2}{a}\;\linf{a}\;0  \quad,\quad \frac{(d^-_a)^2}{b}\;\linf{a}\;0
\end{equation}
and
\begin{equation}
\label{eq:lr15}
q^+\bydef \min_{x\in \LB{d^+_a}a} q_x > \max_{x \in\LB{d^+_a}a} p_x \defby p^+
\quad,\quad
p^-\bydef \min_{x\in \LB{b}{d^-_a}} p_x > \max_{x \in\LB{b}{d^-_a}} q_x \defby q^-\;.
\end{equation}
Indeed, in this case we can take $K_a =d^+_a + \bigl[q^+/(q^+-p^+)\bigr]$ and $K_b =-d^-_a + \bigl[p^-/(p^--q^-)\bigr]$.  On the other hand, $\EE{\Ta{a}{\ze}}\ge d^+_a + a \bigl[\overline q^+/(\overline q^+-\overline p^+)\bigr]$, where $\overline q^+ = \max \{q_x: x\in\LB{d^+_a}a\}$ and $\overline p^+ = \min \{p_x: x\in\LB{d^+_a}a\}$.  An analogous bound holds interchanging $a\leftrightarrow b$, $q^+\leftrightarrow p^-$ and $p^+\leftrightarrow q^-$.  Hence, conditions \reff{eq:lr15} imply the validity of \reff{H} and \reff{Hb}.

For these examples, the observation of~\cite{YcartMart} ---commented below Definition \ref{def:lr0}--- applies, namely the cut-off of the hitting time at zero is equivalent to the cut-off in the original measure-theoretical sense.  

\medskip\par\noindent
\emph{Ehrenfest model.}  It was introduced by Paul and Tatiana Ehrenfest in 1907 to support Botzmann's ideas by showing a simple example where microscopic reversibility leads to what can be interpreted as macroscopic irreversibility. The model consists in $2N$ particles distributed in two urns.  At each unit of time one of the particles is chosen with uniform probability and changed of urn.  The process follows the number of particles in one of the urns; a dynamic that is equivalent to the projection of a random walk on the hypercube $\{-1,1\}^{2N}$.  In this last form, but in continuous time, the cut-off aspects of the model have been studied in~\cite{DiacAsymp}.  The escape behavior has been analyzed in~\cite{BellHarr}, in terms of the initial urn model, and in~\cite{Bingham,BovMetageing} in its random-walk version for continuous time.  

Our framework applies to the initial discrete-time model.  We call $N\to a$ and shift the urn counting by $-a$, so to obtain a model with drift towards zero.  The resulting family $\xa$ has state space $\LB{-a}a$ and transition probabilities
\begin{equation}
p_x=\frac{a-x}{2a} \quad \mbox{and}\quad q_x=\frac{a+x}{2a}\;.
\label{eq:lr16}
\end{equation}
Unlike our previous examples, both transition rates depends on $a$ and have a drift-less limit ($p_x=q_x=1/2$) when $a\to\infty$.  Furthermore, some elementary calculations (see \cite{olivphd}) show that $K_a=K_{-a}=1/\pia0\sim \sqrt a$.  The model has a strong drift, however, because $\EE{\Ta{a}{\ze}}=\EE{\Ta{-a}{\ze}}\sim a\log a$.


\section{Sufficient conditions for cut-off and escape behavior}

In this section we develop the sufficient conditions to be used in the sequel to prove the behavior of the variables in Theorem \ref{theo:0}.  Here we adopt a rather general set-up, not restricted to birth-and-death processes.

\subsection{Cut-off behavior}

The following sufficient condition for a family of random variables $U^{(a)}$ to exhibit cut-off behavior in the sense of Definition \ref{def:lr0} was first pointed out in~\cite{YcartMart}:
\begin{proposition}\label{condsuffcut} A family of random variables $U^{(a)}$ exhibit cut-off at mean times if
\begin{equation}
\lim_{a\to\infty} \var\Bigl{(}\frac{U^{(a)}}{\EE{U^{(a)}}} \Bigr{)} =0\;.
\end{equation}
\end{proposition}
The proof is a direct application of Tchebyshev inequality:  For $Z=1-U^{(a)}/\EE{U^{(a)}}$, we have that
$\pr{|Z|>\varepsilon}\le \EE{Z^2}/\varepsilon = \var(Z)/\varepsilon$.  This simple observation allows the determination of cut-off behavior without a precise estimation of the cut-off times $\EE{U^{(a)}}$.


\subsection{Escape-time behavior}
We adapt the  ``pathwise approach'' introduced in~\cite{CGOV} to a general setting.  We consider Markov chains $\Xa{t}$ on countable alphabets $\alf$, and we distinguish a (``low'') state $\low\in\alf$, (``high'') regions $\high\subset\alf$ and hitting times
\begin{equation}
\Ta{x}{\high}\;\bydef\; \inf _{h\in\high}\inf\{t>0:\x{\low}{t}=h\}\;.
\end{equation}

The mechanism through which the variables $\Ta{x}{\high}$ develop an asymptotically exponential law is the existence of a much shorter time scale for excursions in the direction $\high\to\low$ than vice-versa. Let us be more precise.

\begin{definition}\label{condmeta}
We say that the mean time from $\high$ to $\ell$ is much shorter than that from $\low$ to $\high$, and we write
\begin{equation}
\EE{\Ta\high\low} \;\ll\; \EE{\Ta{\low}{\high}}\;,
\end{equation}
if the following conditions are satisfied:
\begin{enumerate}
\item Different time scales:
\begin{equation}\label{2echel}
\lim_{a\to\infty} \frac{\sup_{x\in\alf}\EE{\Ta{x}{\low}}}{\EE{\Ta{\low}{\high}}}\;=\;0\;.
\end{equation}

\item Uniform integrability of the sequence
\begin{equation}\label{momeps}
S^{(a)}\bydef\frac{\Ta{\low}{\high}}{\E{\Ta{\low}{\high}}} \;.
\end{equation}
This means that for all $\varepsilon>0$ there must exist $L_\epsilon>0$ such that
\begin{equation}
\E{\,S^{(a)}\,;\,S^{(a)}>K}\,<\,\varepsilon\;,\quad\forall\;a\;.
\end{equation}
\end{enumerate}

\end{definition}

These conditions imply escape behavior:
\begin{theorem} \label{thmeta}
Let $\Xa{t}$ be a family of irreducible and positive recurrent Markov chains on a countable alphabet $\alf$.  If $\low\in\alf$ and $\high\subset\alf$ are such that $\EE{\Ta{\high}{\low}} \ll \EE{\Ta{\low}{\high}}$, then the family $\Ta{\low}{\high}$ exhibits escape-time behavior at mean times.
\end{theorem}

Let us first summarize the central idea of the proof.  It amounts to exhibit the asymptotic factorization 
\begin{equation}\label{eq:fact}
\pr{\frac{\Ta\low\high}{\EE{\Ta\low\high}}> s+t}
-\pr{\frac{\Ta\low\high}{\EE{\Ta\low\high}}> s}
\pr{\frac{\Ta\low\high}{\EE{\Ta\low\high}}> t}
\;\linf{a}\;0
\end{equation}
This follows from the existence, by condition \reff{2echel}, of a time scale that is intermediate between the hitting times in both directions.  Indeed,   
\begin{equation}
\scale \;=\; \sqrt{\EE{\Ta\low\high}\,\sup_{x\in\alf} \EE{\Ta{x}\low}}
\end{equation}
is such that 
\begin{equation}\label{eq:rl20}
\beta_a\;\bydef\;\frac{\scale}{\E{\Ta{\low}{\high}}} \linf{a} 0
\end{equation}
and  
\begin{equation}\label{eq:rl25}
\frac{\sup_{x\in\alf} \EE{\Ta{x}\low}}{\scale} \linf{a} 0\;.
\end{equation}
Note that, by Markov inequality, this last equation implies that
\begin{equation}\label{eq:rl21}
\pr{\mbox{walk visits 0 in an interval of length $\scale$}}\; 
\;\linf{a}\; 1\;.
\end{equation}

If a process started at $\low$ has not hit the region $\high$ after $s\,\E{\Ta{\low}{\high}}$ time units,  then, due to \reff{eq:rl21}, it will almost surely fall down to $\ell$ within an interval of length $\scale$.  Once $\ell$ is visited, the process recovers the initial law, by Markovianness.  Therefore,
\begin{eqnarray}
\lefteqn{\pr{\Ta\low\high > (s+t)\, \EE{\Ta\low\high}}\;\approx\;}\nonumber\\
&&\pr{\Ta\low\high > s\, \EE{\Ta\low\high}}
\pr{\Ta\low\high > (t-\scale)\, \EE{\Ta\low\high}}\;.
\end{eqnarray}
As $\scale$ is negligible with respect to $\EE{\Ta\low\high}$ [condition \reff{eq:rl20}], the factorization \reff{eq:fact} follows.
\medskip

This argument is formalized through a number of technical results contained in the following lemmas.
Let 
\begin{equation}
\Ts{s}{\low}\;\bydef\;\inf\Bigr\{t\geq s\,\EE{\Ta\low\high}:\X{(a)}{\low}{t}=\low\Bigr\}\;,
\end{equation}
denote the first return time to $\low$ after $s\,\EE{\Ta\low\high}$, and
\begin{equation}
\Ss{s}{\low}\;\bydef\;\frac{\Ts{s}{\low}}{\EE{\Ta\low\high}}\;.
\end{equation}
These times are well defined: since, as we assume the chain to be positively recurent, every mean hitting time is finite.

\begin{lemma}\label{lem2} Under the hypotheses of Theorem \ref{thmeta}, 
\begin{equation}
\lim_{a\to\infty} \pr{\Ss{s}{\low}>s+\beta_a}=0\;.
\end{equation}
\end{lemma}

\proof
Define $t_a\bydef s\,\E{\Ta\low\high}$, $s_a\bydef \lceil t_a \rceil$, $r_a\bydef\lfloor t_a+\scale\rfloor$ and $I_a\bydef\{s_a,\ldots ,r_a\}$,  
where $\lfloor x\rfloor$ and $\lceil x \rceil$ denote respectively the floor and ceiling functions of $x$.
Partitionning over all possible values of $\X{(a)}{l_a}{s_a}$, we get
\begin{align}
&\pr{\Ss{s}{\low}>s+\beta_a}\nonumber\\
&=\pr{\X{(a)}{\low}{k}\neq \low\,,\,\forall k\in I_a} \nonumber\\
&=\;\sum_{x\in \mathcal{A}_a} \pr{\X{(a)}{\low}{k}\neq \low\,,\,\forall k\in I_a \Bigm{|} \X{(a)}{\low}{s_a}=x} \,\pr{\X{(a)}{\low}{s_a}=x}\\
&\leq\; \sup_{x\in \mathcal{A}_a} \pr{\X{(a)}{\low}{k}\neq \low \,,\,\forall k\in I_a \Bigm{|} \X{(a)}{\low}{s_a}=x}\, \sum_{x\in \mathcal{A}_a} \pr{\X{(a)}{\low}{s_a}=x} \nonumber\\
&=\;\sup_{x\in \mathcal{A}_a} \pr{\Ta{x}{\low}>\scale} \;,\nonumber
\end{align}
Hence, using Markov inequality,
\begin{equation}
\pr{\Ss{s}{\low}>s+\beta_a}\;\leq\; \frac{\sup_{x\in\alf}\EE{\Ta{x}{\low}}}{\scale}\;,
\end{equation}
which tends to zero due to \reff{2echel}. \qed
\medskip

\begin{lemma}\label{lem3} Under the hypotheses of Theorem \ref{thmeta}, the following inequalities hold:
\begin{equation}\label{eq:rl40}
\pr{S^{(a)}>s+t\,,\, \Ss{s}{\low}\leq s+\beta_a} \;\geq\; \pr{S^{(a)}>s+\beta_a \,,\, \Ss{s}{\low}\leq s+\beta_a}\, \pr{S^{(a)}>t} 
\end{equation}
and
\begin{equation}\label{eq:rl41}
\pr{S^{(a)}>s+t\,,\, \Ss{s}{l_a}\leq s+\beta_a} \;\leq\; \pr{S^{(a)}>s}\,\pr{S^{(a)}> t-\beta_a}\;.
\end{equation}
\end{lemma}

\proof
We start with the following identity, obtained by partionning over all possible values of $\Ts{s}{\low}$, 
\begin{eqnarray}\label{partit}
\lefteqn{\hspace{-1cm}\pr{S^{(a)}>s+t\,,\, \Ss{s}{\low}\leq s+\beta_a}} \nonumber\\
&=&\sum_{k\in I_a}\,\pr{S^{(a)}>s+t\,,\, \Ts{s}{\low}=k } \nonumber\\
		&=&\sum_{k\in I_a}\,\pr{S^{(a)}>s+t\,,\, S^{(a)}>s  \,,\, \Ts{s}{\low}=k } \nonumber\\
		&=&\sum_{k\in I_a}\,\pr{S^{(a)}>s+t\,\Bigm{|}\,S^{(a)}>s\,,\,\Ts{s}{\low}=k } \;\pr{S^{(a)}>s\,,\,\Ts{s}{\low}=k} \;. 
\end{eqnarray}
The Markov property implies that
\begin{equation}\label{mkvlem3}
\pr{S^{(a)}>s+t\,\Bigm{|}\,S^{(a)}>s\,,\,\Ts{s}{\low}=k }\;=\;\pr{S^{(a)}>s+t-\frac{k}{\EE{\Ta\low\high}} } \;,
\end{equation}
for all $k\in I_a$.  The rightmost expression is bounded below by its value at $k=s_a$, and bounded above by its value at $k=s_a+\scale$.  Hence,
\begin{equation}
\label{eq:rl30}
\pr{S^{(a)}>t} \;\le\; 
\pr{S^{(a)}>s+t\,\Bigm{|}\,S^{(a)}>s\,,\,\Ts{s}{\low}=k } \;\le\;
\pr{S^{(a)}>t-\beta_a}\;.
\end{equation}

To prove \reff{eq:rl40} we combine \reff{partit} with the leftmost bound in \reff{eq:rl30}.  We obtain:
\begin{align}
\pr{S^{(a)}>s+t\,,\, \Ss{s}{\low}\leq s+\beta_a} &\geq\; \pr{S^{(a)}>t}\;\sum_{k\in I_a}\,\pr{S^{(a)}>s\,,\,\Ts{s}{\low}=k} \nonumber\\
		&=\;\pr{S^{(a)}>t}\;\pr{S^{(a)}>s \,,\, \Ss{s}{\low}\leq s+\beta_a}  \\    
		&\geq\; \pr{S^{(a)}>t}\;\pr{S^{(a)}>s+\beta_a \,,\, \Ss{s}{\low}\leq s+\beta_a} \;. \nonumber
\end{align}

Likewise, \reff{eq:rl41} follows from \reff{partit}, the rightmost bound in \reff{eq:rl30}:
\begin{align}
\pr{S^{(a)}>s+t\,,\, \Ss{s}{\low}\leq s+\beta_a} &\leq\; \pr{S^{(a)}>t-\beta_a}
\;\sum_{k\in I_a}\,\pr{S^{(a)}>s\,,\,\Ts{s}{\low}=k} \\
&\leq\; \pr{S^{(a)}>t-\beta_a}\,\pr{S^{(a)}>s}\;.\qed \nonumber
\end{align}
 
\begin{lemma}\label{lem5}
Assume the hypothesis of Theorem \ref{thmeta} and let $S$ be the weak limit of a subsequence of $S^{(a)}$.  Then, 
\begin{equation}\label{eq:rl49}
\pr{S>s+t}\;=\;\pr{S>s}\, \pr{S>t}
\end{equation}
for all $s$ and $t$ such that $s$, $t$ and $s+t$ are continuity points of the distribution function of $S$.
\end{lemma}
\proof The previous lemmas remain valid for $a$ varying through the subsequence index.  For notational simplicity we keep using $a$ for the latter and denote $S^{(a)}$ the converging subsequence.  Identity \reff{eq:rl49} is proven by decomposing it into an upper and a lower bound.
\medskip

\noindent \emph{Proof of the upper bound.}
We write:
\begin{equation}\label{eq:rl50}
\pr{S^{(a)}>s+t}- \pr{S^{(a)}>s}\,\pr{S^{(a)}>t}\;=\; A_a + B_a
\end{equation}
with
\begin{equation}\label{eq:rl51}
A_a\;=\;\pr{S^{(a)}>s+t}-\pr{S^{(a)}>s}\,\pr{S^{(a)}>t-\beta_a}
\end{equation}
and
\begin{equation}\label{eq:rl52}
B_a\;=\; \pr{S^{(a)}>s} \Bigl{[} \pr{S^{(a)}>t-\beta_a}- \pr{S^{(a)}>t}\Bigr{]}\;. 
\end{equation}
By \reff{eq:rl41},\\
\begin{eqnarray}\label{eq:rl53}
A_a &\le & \pr{S^{(a)}>s+t} - \pr{S^{(a)}>s+t\,,\, \Ss{s}{\low}\leq s+\beta_a}\nonumber\\
&=& \pr{S^{(a)}>s+t\,,\, \Ss{s}{\low}> s+\beta_a}\;.
\end{eqnarray}
Thus, by Lemma \ref{lem2}, 
\begin{equation}
\label{eq:rl54}
\limsup_{a\to\infty} A_a \;\le\; 0\;.
\end{equation}
On  the other hand, as $t$ is a point of continuity and $\beta_a\to 0$ [condition \reff{eq:rl20}], the weak convergence of the sequence implies that
\begin{equation}
\lim_{a\to\infty} \Bigl[\pr{S^{(a)}>t-\beta_a}- \pr{S^{(a)}>t}\Bigr]\;=\;0\;.
\end{equation}
Thus
\begin{equation}\label{eq:rl56}
\lim_{a\to\infty} B_a\;=\; 0\;.
\end{equation}
From \reff{eq:rl50}, \reff{eq:rl54}, \reff{eq:rl56} and the weak convergence of $S^{(a)}$ we conclude that
\begin{equation}
 \pr{S>s+t}-\pr{S>s}\, \pr{S>t} \;\le\; 0
\end{equation}

\noindent \emph{Proof of the lower bound.}
This time we write
\begin{equation}\label{eq:rl60}
\pr{S^{(a)}>s+t}- \pr{S^{(a)}>s}\, \pr{S^{(a)}>t} \;=\; C_a + D_a
\end{equation}
with
\begin{equation}\label{eq:rl61}
C_a\;=\; \pr{S^{(a)}>s+t}-\pr{S^{(a)}>s+\beta_a\,,\,\Ss{s}{\low}\leq s+\beta_a}\,\pr{S^{(a)}>t}
\end{equation}
and
\begin{equation}\label{eq:rl62}
D_a\;=\;
\pr{S^{(a)}>t}\Bigl{[}\pr{S^{(a)}>s+\beta_a\,,\,\Ss{s}{\low}\leq s+\beta_a}- \pr{S^{(a)}>s}\Bigr{]}\;. 
\end{equation}
By \reff{eq:rl40}, 
\begin{equation}\label{eq:rl63}
C_a \;\ge\;\pr{S^{(a)}>s+t\,,\,\Ss{s}{\low}> s+\beta_a}\;\linf{a} 0
\end{equation} 
where the convergence is due to Lemma \ref{lem2}.
We further decompose $D_a$ by substracting and adding $\pr{S^{(a)}>s+\beta_a}$.  We obtain
\begin{equation}\label{eq:rl64}
D_a\;=\;
-\pr{S^{(a)}>s+\beta_a,\Ss{s}{\low}>s+\beta_a}+\Bigl[\pr{S^{(a)}>s+\beta_a}-\pr{S^{(a)}>s}\Bigr]\;. 
\end{equation}
Both terms converge to zero; the first one by Lemma \ref{lem2} and the second by continuity at $s$ and condition \reff{eq:rl20}.  Therefore,
\begin{equation}
\label{eq:rl65}
\lim_{a\to\infty} D_a \;=\; 0\;.
\end{equation}
Because of the weak convergence of $S^{(a)}$, equations \reff{eq:rl60}, \reff{eq:rl63} and \reff{eq:rl65} imply that
\begin{equation}
 \pr{S>s+t}-\pr{S>s}\, \pr{S>t} \;\ge\; 0\;.\qed
\end{equation}

\paragraph{Proof of Theorem \protect\ref{thmeta}.}
It follow from the preceding lemma and some general considerations.  
We first remark that the sequence $S^{(a)}$ is uniformly tight because $\E{S^{(a)}}=1$.  Indeed, using Markov inequality,
\begin{equation}
\pr{S^{(a)}>t}\;\leq\; \,\frac{\E{S^{(a)}}}{t}\;=\;\frac{1}{t} \;\linf{t} \;0\;.
\end{equation}
It follows (see, e.g., Theorem. 4.4.3. in \cite{Chung}) that $S^{(a)}$ admits a sub-sequence $S^{(a')}$ that converges in law.  We only have to prove that there exists a unique limit law, namely ${\mathcal{E}xp}(1)$.

Let $S$ be the limit of one of these subsequences .  By Lemma \ref{lem5}
\begin{equation}\label{caractexp}
\pr{S>s+t}=\pr{S>s}\, \pr{S>t}\;,
\end{equation}
for all continuity points of the distribution function $F_S$ of $S$. These discontinuity points are dense in $\mathbb{R}$ because $F_s$, being increasing, has at most a denumerable number of discontinuities.  Thus, \reff{caractexp} holds for all $s$ and $t$.

It remains to show that $\E{S}=1$ to prove that $S$ can have only one law, namely exponential with mean 1. It is here that we invoke the hypothesis of uniform integrability of the sequence $S^{(a)}$.  Indeed,  
by Theorem 5.4 of \cite{Billcvce1}, the fact that $S^{(a')}$ is a weakly convergent sequence of uniformly integrable mean-one random variables, implies that 
\begin{equation}
\E{S}=\lim_{a\to\infty}\,\EE{S^{(a'')}}=1 \;. \qed
\end{equation}


\section{Mean hitting times for birth-and-death processes}\label{Htim}

The estimation needed to apply the criteria of the previous section to birth-and-death processes relies on explicit formulas of mean hitting times available for these processes.  Throughout this section, to shorten formulas we adopt the conventions $\prod_{j+1}^j\bydef 1$ and $\sum_a^{a-1}\bydef 0$.  

We consider an irreducible birth-and-death process $X$ on a finite interval $\LB{b}{a} \subset\mathbb{Z}$ with transition rates $0<p(x,x+1)\defby p_x$, $b\le x\le a-1$; $0<p(x,x-1)\defby q_x$, $b+1\le x\le a$, and $r_x\bydef 1-p_x-q_x$.  The invariant measure for such a process is also reversible and can be written in the forms [c.f.\ \reff{invmes}]:
 \begin{equation}\label{invmes-bis}
\pia{x}\;=\;
\prod_{i=b+1}^{x}\, \frac{p_{i\mo}}{q_i} \,\pia{b} \;=\;
\prod_{i=x+1}^{a}\, \frac{q_{i}}{p_{i\mo}} \,\pia{a} 
\end{equation}
for $b\le x\le a$.  Hence, for $k<l$, 
\begin{equation}\label{pipi}
\prod_{i=k+1}^{l} \, \frac{p_{i\mbox{\tiny\em-1}}}{q_i} \;=\; \prod_{i=k+1}^{l} \, \frac{q_i}{p_{i\mbox{\tiny\em-1}}} \;=\; \frac{\pia{l}}{\pia{k}}\;.
\end{equation}
\begin{proposition}\label{MomenTrans}
For the preceding birth-and-death process, the following identities hold: 
\begin{enumerate}
\item For $b\le n < j \le a$, 
	\begin{eqnarray}\label{eq:rl70}
\EE{\T jn} &=& \sum_{k=n+1}^j \frac{\pi(\LB ka)}{q_k\,\pi(k)}\\
\label{eq:rl73.1}
\E{\T{j}{n}^2}&=&\sum_{k=n+1}^{j} \,\frac{2}{q_k\,\pi(k) } \, \sum_{l=k}^{a} \, \E{\T{l}{n}}\,\pi(l)\; -\; \E{\T{j}{n}}\;.
	\end{eqnarray}
\item For $b\le j < n \le a$, 
\begin{eqnarray}\label{eq:rl71}
\EE{\T jn} &=& \sum_{k=j}^{n-1} \frac{\pi(\LB bk)}{p_k\,\pi(k)} \\
\label{eq:rl73}
\E{\T{j}{n}^2}&=&\sum_{k=j}^{n-1} \,\frac{2}{p_k\,\pi(k) } \, \sum_{l=b}^{k} \, \E{\T{l}{n}}\,\pi(l)\; -\; \E{\T{j}{n}}\\
\label{eq:rl72}
\EE{\T jn} + \EE{\T nj} &=& \sum_{k=j+1}^n \frac{1}{q_k\,\pi(k)}\;.
	\end{eqnarray}
\item For $b\le m< j < n \le a$, 
	\begin{eqnarray}\label{espappend+-}
\E{\T{j}{\{m,n\}}} &=&	\frac{\E{\T{n}{m}}\,\E{\T{j}{n}}-\E{\T{m}{n}}\,\E{\T{n}{j}}}{\E{\T{m}{n}}+\E{\T{n}{m}}}\\[5pt]
\label{espappend+-1}
&=&	\frac{\E{\T{m}{n}}\,\E{\T{j}{m}}-\E{\T{n}{m}}\,\E{\T{m}{j}}}{\E{\T{m}{n}}+\E{\T{n}{m}}}\;.
	\end{eqnarray}
\end{enumerate}
\end{proposition}

\proof 
We apply the standard technique of establishing a difference equation for consecutive mean times that is solved by iteration (see for instance \cite{Feller1} chapter XIV).  In all cases, the equation is obtained from the following decomposition.
\medskip\par\noindent
\emph{Claim:} Let $\med\subset \LB ab$ and $j\not\in\med$, $b\le j \le a$.  Then, for any function $F$,
\begin{equation}
\label{eq:claim}
\EE{F(\T{j}{\med})} \;=\; p_j\,\EE{F(\T{j+1}{\med}+1)} + q_j \,\EE{F(\T{j-1}{\med}+1)} +r_j \,\EE{F(\T{j}{\med}+1)}\;.
\end{equation}
with the conventions $\T{a+1}{\med}, \T{b-1}{\med} \bydef 0$ and $, \T{i}{\med}\bydef 0$ if $i\in\med$.
\medskip\par\noindent
Indeed, decomposing according to the first step of the process, 
\begin{eqnarray}
\label{eq:claim1}
\EE{F(\T{j}{\med})} &=& p_j\,\EEE{F(\T{j}{\med})\Bigm|X(1)=j+1} 
+ q_j \,\EEE{F(\T{j}{\med})\Bigm|X(1)=j-1} \nonumber\\
&&  \qquad\qquad {}+r_j \,\EEE{F(\T{j}{\med}+1)\Bigm|X(1)=j}\;.
\end{eqnarray}
In the three cases $i=j-1,j,j+1$,
\begin{eqnarray}
\label{eq:claim2}
\EEE{F(\T{j}{\med})\Bigm|X(1)=i} &=& \sum_{k\ge 1} F(k)\, \prr{\T{j}\med=k\Bigm|X(1)=i}\nonumber\\
&=& \sum_{k\ge 1} F(k)\, \pr{\T{i}{\med}=k-1}\nonumber\\
&=& \sum_{k\ge 0} F(k+1)\, \pr{\T{i}{\med}=k}\nonumber\\
&=& \EEE{F(\T{i}{\med}+1)}\;.
\end{eqnarray}
The second equality is due to Markovianness.  The claim follows from \reff{eq:claim1} and \reff{eq:claim2}.  

\bigskip\par\noindent
\emph{Proof of \reff{eq:rl70} and \reff{eq:rl71}.}
Applying \reff{eq:claim} for $F(T)=T$ and $\med =\{n\}$ we arrive to the difference equation
\begin{equation}
\label{eq:dif1}
\EE{\T jn} \;=\; \frac{p_j}{p_j+q_j} \EE{\T{j+1}n} + \frac{q_j}{p_j+q_j} \EE{\T{j-1}n}
+ \frac{1}{p_j+q_j} 
\end{equation}
valid for $j=b+1,\ldots,a-1$ and subject to the boundary conditions:
\begin{eqnarray}\label{eq:dif2}
\EE{\T bn} &=&\EE{\T{b+1}n} + \frac1{p_b} \nonumber\\
 \EE{\T nn} &=&0\\
\EE{\T an} &=& \EE{\T{a-1}n} + \frac1{q_a} \;.\nonumber
\end{eqnarray}

To prove \reff{eq:rl70} it is useful to rewrite \reff{eq:dif1} in the form
\begin{equation}
\label{eq:dif3}
D_{k} - D_{k-1} \;=\; B_k(D_{k+1}-D_{k}) + \epsilon_k
\end{equation}
with $D_j=\EE{\T jn}$, $B_j=p_j/q_j$ and $\epsilon_j=1/q_j$.  This equation is easily solved by iteration, yielding
\begin{equation}
\label{eq:dif4}
D_{k} - D_{k-1} \;=\; \prod_{i=k}^{a-1} B_i (D_a-D_{a-1}) + \sum_{l=k}^{a-1} \epsilon_l \prod_{i=k}^{l-1} B_i\;,
\end{equation}
and, thus, 
\begin{equation}
\label{eq:dif5}
D_j \;=\; D_a-\sum_{k=j+1}^{a}\Bigl(\prod_{i=k}^{a-1} B_i \,(D_a-D_{a-1}) + \sum_{l=k}^{a-1} \epsilon_l \prod_{i=k}^{l-1} B_i\Bigr)\;.
\end{equation}
Upon application of the last two conditions in \reff{eq:dif2}:
\begin{eqnarray}\label{eq:dif6}
D_a-D_{a-1} &=& \epsilon_a\nonumber\\
D_n &=& 0\;,
\end{eqnarray}
we finally obtain
\begin{equation}
\label{eq:dif5b}
D_j \;=\; \sum_{k=n+1}^{j} \sum_{l=k}^{a} \epsilon_l \prod_{i=k}^{l-1} B_i\;,
\end{equation}
that is, 
\begin{equation}\label{espappend-}
\E{\T{j}{n}}\;=\;\sum_{k=n+1}^{j} \,\frac{1}{q_k} \, \sum_{l=k}^{a} \, \prod_{i=k+1}^{l} \, \frac{p_{i\mbox{\tiny\em-1}}}{q_i}\;.
\end{equation}
This is, in fact, expression \reff{eq:rl70} due to identity \reff{pipi}.

The proof of \reff{eq:rl71} is analogous, but inverting the direction of iteration.  We write \reff{eq:dif1} in the form 
\begin{equation}
\label{eq:dif7}
D_{k+1} - D_k \;=\; \widetilde B_k\,(D_k-D_{k-1}) - \widetilde \epsilon_k
\end{equation}
with $\widetilde B_j=q_j/p_j$ and $\widetilde \epsilon_j=1/p_j$.  The solution is now 
\begin{equation}
\label{eq:dif8}
D_j \;=\; D_b+\sum_{k=b}^{j-1}\Bigl(\prod_{i=b+1}^k \widetilde B_i \,(D_{b+1}-D_{b}) - \sum_{l=b+1}^k \widetilde \epsilon_l \prod_{i=l+1}^k \widetilde B_i\Bigr)\;.
\end{equation}
The first two conditions in \reff{eq:dif2}:
\begin{eqnarray}\label{eq:dif9}
D_{b+1}-D_b &=& -\widetilde\epsilon_b\nonumber\\
D_n &=& 0\;.
\end{eqnarray}
lead to
\begin{equation}
\label{eq:dif8b}
D_j \;=\sum_{k=j}^{n-1} \sum_{l=b}^k \widetilde \epsilon_l \prod_{i=l+1}^k \widetilde B_i\;.
\end{equation}
In terms of the original quantities this is equal to
\begin{equation}\label{espappend+}
\E{\T{j}{n}}\;=\;\sum_{k=j}^{n-1} \,\frac{1}{p_k} \,  \sum_{l=0}^{k} \, \prod_{i=l+1}^{k} \, \frac{q_i}{p_{i\mbox{\tiny\em-1}}}\;,
\end{equation}
which, together with \reff{pipi}, leads to \reff{eq:rl71}.

\bigskip\par\noindent
\emph{Proof of \reff{eq:rl73.1} and \reff{eq:rl73}.}
We apply \reff{eq:claim} for $F(T)=T^2$ and $\med=\{n\}$ to arrive to the difference equation
\begin{equation}
\label{eq:dif10}
\EE{\T jn^2} \;=\; \frac{p_j}{p_j+q_j} \EE{\T{j+1}n^2} + \frac{q_j}{p_j+q_j} \EE{\T{j-1}n^2}
+ \frac{2\EE{\T jn} -1}{p_j+q_j} \;,
\end{equation}
$j\in\LB{b+1}{a-1}$, plus the boundary conditions:
\begin{eqnarray}\label{eq:dif12}
\EE{\T bn^2} &=&\EE{\T{b+1}n^2} + \frac{2\EE{\T bn} -1}{p_b} \nonumber\\
 \EE{\T nn^2} &=&0\\
\EE{\T an^2} &=& \EE{\T{a-1}n^2} + \frac{2\EE{\T an} -1}{q_a} \;.\nonumber
\end{eqnarray}
If we now substitute $D_j=\EE{\T jn^2}$, $B_j=p_j/q_j$, $\widetilde B_j=q_j/p_j$,
$\epsilon_j=\bigl(2\EE{\T jn}-1\bigr)/q_j$ and $\widetilde \epsilon_j=\bigl(2\EE{\T jn}-1\bigr)/p_j$
we recover equations \reff{eq:dif3} and \reff{eq:dif7} and the boundary conditions \reff{eq:dif6} and \reff{eq:dif9}.  The corresponding solutions \reff{eq:dif5b} and \reff{eq:dif8b} yield
\begin{equation}\label{eq:tired}
\E{\T{j}{n}^2} \;=\; \left\{\begin{array}{rl}
\displaystyle\sum_{k=n+1}^{j} \,\frac{1}{q_k } \, \sum_{l=k}^{a} \, \Bigl(2\E{\T{l}{n}}-1\Bigr) \prod_{i=k+1}^{l} \, \frac{p_{i\mbox{\tiny\em-1}}}{q_i} & b\le n<j\le a\\[15pt]
\displaystyle\sum_{k=j}^{n+1} \,\frac{1}{p_k } \, \sum_{l=b}^{k} \, \Bigl(2\E{\T{l}{n}}-1\Bigr)  \, \prod_{i=l+1}^{k} \, \frac{q_i}{p_{i\mbox{\tiny\em-1}}} & b\le j<n \le a\;.
\end{array}\right.
\end{equation}
The first and second lines become, respectively, \reff{eq:rl73.1} and \reff{eq:rl73} upon resorting to \reff{pipi} and \reff{espappend-}/\reff{espappend+}.

\bigskip\par\noindent
\emph{Proof of \reff{eq:rl72}.}
From \reff{eq:rl70}, \reff{eq:rl71} and the reversibility relation $\pi(x)\,p_x=\pi(x+1)\, q_{x+1}$, we have:
\begin{equation}\label{eq:rl70+}
\EE{\T jn} + \EE{\T nj} \;=\; \sum_{k=j+1}^n \frac{\pi(\LB b{k-1}) + \pi(\LB ka)}{q_k\,\pi(k)}\,
\end{equation}
which is, precisely, \reff{eq:rl72} because $\pi(\LB ba)=1$.

\bigskip\par\noindent
\emph{Proof of \reff{espappend+-} and \reff{espappend+-1}.}
The application of  \reff{eq:claim} for $F(T)=T$ and $\med =\{m,n\}$ leads to the difference equation \reff{eq:dif1} with $n$ replaced by $\{m,n\}$.  The equation is valid for $m<j<n$ and is supplemented by the boundary conditions
\begin{equation}\label{eq:dif2bis}
 \EE{\T nn} \;=\;0\;=\; \EE{\T mm}\;.
\end{equation}
The substitutions $D_j=\EE{\T j{\{m,n\}}}$, $B_j=p_j/q_j$ and $\epsilon_j=1/q_j$ lead to the same identity \reff{eq:dif3} which, iterated from $k=n$ downwards, yields
\begin{equation}
\label{eq:difn1}
D_j \;=\; D_n-\sum_{k=j+1}^{n}\Bigl(\prod_{i=k}^{n-1} B_i \,(D_n-D_{n-1}) + \sum_{l=k}^{n-1} \epsilon_l \prod_{i=k}^{l-1} B_i\Bigr)\;.
\end{equation}
Once imposed the condition $D_n=0$, the condition, the solution takes the form
  \begin{equation}
\label{eq:difn2}
D_j\;=\; D_{n-1}\,F_j - G_j\;,
\end{equation}
with
\begin{eqnarray}\label{eq:difn3}
F_j &=& \sum_{k=j+1}^n \prod_{i=k}^{n-1} B_i\;,\\
\label{eq:difn4}
G_j &=& \sum_{k=j+1}^n \sum_{l=k}^{n-1} \epsilon_l \prod_{i=k}^{l-1} B_i\;.
\end{eqnarray}
The remaining condition $D_m=0$ implies, then,
 \begin{equation}
\label{eq:difn5}
D_j\;=\; \frac{G_m}{F_m}\,\,F_j - G_j\;.
\end{equation}
Replacing $B_i=p_i/q_i$, $\epsilon_j=1/q_j$ and using \reff{pipi} we get 
\begin{equation}\label{eq:difn6}
F_j \;=\; \sum_{k=j+1}^n  \frac{q_n\,\pi(n)}{q_k\,\pi(k)}
\end{equation}
and
\begin{eqnarray}\label{eq:difn7}
G_j &=& \sum_{k=j+1}^n \frac{\pi(\LB k{n-1})}{q_k\,\pi(k)}\nonumber\\
&=& \sum_{k=j+1}^n \frac{\pi(\LB ka)-\pi(\LB na)}{q_k\,\pi(k)}\;.
\end{eqnarray}
We now resort to \reff{eq:rl70} and \reff{eq:rl72} to conclude:
\begin{eqnarray}\label{eq:difn8}
F_j &=& q_n\,\pi(n) \bigl[ \E{\T jn} + \E{\T nj} \bigr]\;,\\[5pt]
\label{eq:difn9}
G_j &=& \E{\T nj} - \pi(\LB na) \bigl[ \E{\T jn} + \E{\T nj} \bigr]\;.
\end{eqnarray}
Identity \reff{espappend+-} is obtained by replacing these two equalities (and their $j=m$ version) into \reff{eq:difn5}.  Identity \reff{espappend+-1} can, of course, be obtained in a similar fashion iterating \reff{eq:dif7} up from $k=m$.  Alternatively, it follows from the fact that the numerators of \reff{espappend+-} and \reff{espappend+-1} are equal because their difference is 
\begin{eqnarray}\label{eq:difn11}
\lefteqn{\E{\T{n}{m}}  \bigl(\E{\T{m}{j}}+\E{\T{j}{n}} \bigr) - \E{\T{m}{n}} \bigl(\E{\T{n}{j}} + \E{\T{j}{m}}\bigr)}\nonumber\\[5pt]
&&=\; \E{\T{n}{m}} \,\E{\T{m}{n}} - \E{\T{m}{n}} \,  \E{\T{n}{m}}  \;=\; 0\;. \qed
\end{eqnarray}
\bigskip

Next we list two useful consequences of the previous proposition.
\begin{corollary}  Consider the process of Proposition \ref{MomenTrans}.  Then, for $b+1\le x \le a$,
\begin{equation}
\label{eq:rl74}
\EE{\T{x}{x-1}^2} \;=\; \frac{2}{q_x\,\pi(x)} \sum_{c=x}^a \frac{\pi\bigl(\LB xa\bigr)^2}{q_c\,\pi(c)}\; -\; \EE{\T{x}{x-1}}\;,
\end{equation}
and, for $b\le j<n \le a$,
\begin{equation}
\label{eq:rlgl}
\EE{\T{j}{n}^2} \;\le\; 2\, \EE{\T{j}{n}}^2 + 2\, \EE{\T{b}{j}}\,\EE{\T{j}{n}} - \EE{\T{j}{n}}\;.
\end{equation}
\end{corollary}
\proof Display \reff{eq:rl74} is an immediate consequence of \reff{eq:rl73.1} and \reff{eq:rl70}.  Inequality \reff{eq:rlgl} is obtained from \reff{eq:rl73} and \reff{eq:rl71}:
\begin{eqnarray}\label{eq:rlgl1}
\E{\T{j}{n}^2}&\le&\sum_{k=j}^{n-1} \,\frac{2}{p_k\,\pi(k) } \, \sum_{l=b}^{k} \bigl(\E{\T lj}+\E{\T{j}{n}}\bigr)\,\pi(l)\; -\; \E{\T{j}{n}}\nonumber\\
&=& \sum_{k=j}^{n-1} \,\frac{2}{p_k\,\pi(k) } \, \sum_{l=b}^{k} \E{\T lj}\,\pi(l) + 2 \,\E{\T{j}{n}}^2 -\; \E{\T{j}{n}}\nonumber\\
&\le& 2\,\E{\T bj} \sum_{k=j}^{n-1} \,\frac{\pi(\LB bk)}{p_k\,\pi(k) } + 2 \,\E{\T{j}{n}}^2 -\; \E{\T{j}{n}}\nonumber\\
&=& 2\, \EE{\T{b}{j}}\,\EE{\T{j}{n}} + 2\, \EE{\T{j}{n}}^2 - \EE{\T{j}{n}}\;.\qed
\end{eqnarray}

\section{Cut-off and escape for half-well models}\label{HWmodel}

We shall proof parts 2 and 3 of Theorem \ref{theo:0} in two stages.  In this section we consider the half-well models, obtained when setting $b=0$ and $q_0=0$ or $a=0$ and $p_0=0$.  The actual result is a little stronger than the statement of Theorem \ref{theo:0}.  It is based on the quantities
\begin{equation}
\qa(x) \bydef \frac{1}{\pia{\Bx}} \sum_{c\in\LB xa} \frac{\pia{\LB ca}^2}{q_c\,\pia c}
\quad;\quad \qa \bydef \sup_{x\in\LB0a} \qa(x)\;.
\label{eq:rl80}
\end{equation}
As
\begin{equation}
\pia{\LB1a}\,\qa(1)\;\le\qa(1)\;\le\;\qa \;\le\; K_a^2\,K_q\;,
\label{eq:rl81}
\end{equation}
the following lemma implies parts 2 and 3 of Theorem \ref{theo:0} 
\begin{lemma}\label{lem:cohw}
Let $\xa$ be a family of irreducible birth-and-death chains on $\LB{0}{a}$.
\begin{itemize}
\item [(i)] If 
\begin{equation} \label{eq:rl82.1}
\pia{ \LB1a}\,\frac{\qa(1)}{\EE{\Ta{a}{\ze}}}\linf{a}0 \;,
\end{equation}
then $\EE{\Ta a\ze}\ll\EE{\Ta\ze a}$ and, as a consequence, the random variables $\Ta{\mbox{\tiny\em0}}{a}$ have escape-time behavior at mean times.
\item [(ii)] If 
\begin{equation} \label{eq:rl82}
 \frac{\qa}{\EE{\Ta{a}{\ze}}}\linf{a}0 \;,
\end{equation}
then $\var \bigl(\Ta{a}{\mbox{\tiny\em0}}/\EE{\Ta{a}{\mbox{\tiny\em0}}}\bigr)\longrightarrow 0$ as $a\to\infty$ and, as a consequence, the random variables
$\Ta{a}{\mbox{\tiny\em0}}$ have cut-off behavior at mean times.
\end{itemize}
\end{lemma}
\proof

\emph{(i)} 
We have to verify that both conditions in Definition \ref{condmeta} are valid for $\low=0$, $\high = \{a\}$; escape behavior follows then from Theorem \ref{thmeta}.  The second condition is a consequence of \reff{eq:rlgl} which for the half well implies that
\begin{equation} \label{eq:rl85}
\EE{(\Ta\ze a)^2} \;\le\; 2\,\EE{\Ta \ze a}^2\;.
\end{equation}
Therefore the sequence $S^{(a)} = \Ta\ze a/\EE{\Ta\ze a}$ is uniformly square integrable and thus, by Schwartz inequality, uniformly integrable.
\medskip

The first condition amounts to proving that $\EE{\Ta a\ze}/\EE{\Ta \ze{a}} \to 0$ as $a\to\infty$ or, equivalently, that 
\begin{equation} \label{eq:rl86}
\frac{\EE{\Ta a\ze}}{\EE{\Ta a\ze} + \EE{\Ta\ze a}} \;\linf a0\;.
\end{equation}
By \reff{eq:rl70} and \reff{eq:rl72} this corresponds to 
\begin{equation} \label{eq:rl87}
\Gamma^{(a)}\;\bydef\; \frac{\displaystyle\sum_{x=1}^a \frac{\pia{\Bx}}{q_x\,\pia x}}{\displaystyle\sum_{x=1}^a \frac{1}{q_x\,\pia x}}
\;\linf a0\;.
\end{equation}
To prove this we consider the probability measure on $\LB0a$ defined by expectations
\begin{equation} \label{eq:rl88}
\mathcal{E}(F) \;\bydef\; \frac{\displaystyle\sum_{x=1}^a \frac{F(x)}{q_x\,\pia x}}{\displaystyle\sum_{x=1}^a \frac{1}{q_x\,\pia x}}\;.
\end{equation}
Indeed, the inequality $\mathcal{E}\bigl(F\bigr)^2\le \mathcal{E}\bigl(F^2\bigr) $ for $F(x)=\pia{\Bx}$ implies that l
\begin{equation} \label{eq:rl89}
\Gamma^{(a)} \;\le\; \frac{\displaystyle\sum_{x=1}^a \frac{\pia{\Bx}}{q_x\,\pia x}^2}{\displaystyle\sum_{x=1}^a \frac{\pia{\Bx}}{q_x\,\pia x}} \;=\;
\frac{\displaystyle\sum_{x=1}^a \frac{\pia{\Bx}}{q_x\,\pia x}^2}{\EE{\Ta a\ze}}\;.
\end{equation}
Thus, 
\begin{equation} \label{eq:rl90}
\Gamma^{(a)} \;\le\; \pia{\LB1a} \,\frac{\qa(1)}{\EE{\Ta a\ze}} \;,
\end{equation}
and \reff{eq:rl87} is a consequence of hypothesis \reff{eq:rl82.1}.

\medskip\par\noindent
\emph{(ii)} 
Due to the independence of the hitting times and to \reff{eq:rl74},
\begin{eqnarray}\label{eq:rl83}
\var\bigl{(}\Ta{a}{\ze}\bigr{)}
&=&\sum_{x=1}^a \,\var\bigl{(}\Ta{x}{x\mo}\bigr{)} \nonumber\\
&\le& \sum_{x=1}^a \frac{2}{q_x\,\pi(x)} \sum_{c=x}^a \frac{\pi\bigl(\LB ca\bigr)^2}{q_c\,\pi(c)}\;.
\end{eqnarray}
Hence, by \reff{eq:rl70},
\begin{equation} \label{eq:rl84}
\var\Bigl(\frac{\Ta{a}{\ze}}{\EE{\Ta a\ze}}\Bigr)
\;\le\; \frac{\qa}{\EE{\T a0}} \;\linf a0\;.
\end{equation}
The cut-off behavior is, thus, a consequence of Proposition \ref{condsuffcut}.\qed
\section{Cut-off and escape behavior for the full well}\label{sec:full}
\subsection{Comparison (in)equalities}
Part of the results will be obtained by comparison with the half-well hitting times.  Let us therefore define, for each process $\xa$, its left $\xal$ and right $\xar$ half-well versions.  The latter is a birth-and-death process on $\LB 0a$ with transition rates $(\ppr_x,\qqr_x,\rrr_x)=(p_x,q_x,r_x)$ for $1\le x\le a$ and $(\ppr_0,\qqr_0,\rrr_0)=(p_0,0,q_0+r_0)$.  Likewise, $\xal$ is a process on $\LB b0$ with transition rates $(\ppl_x,\qql_x,\rrl_x)=(p_x,q_x,r_x)$ for $b\le x\le -1$ and $(\ppl_0,\qql_0,\rrl_0)=(0,q_0,p_0+r_0)$. From \reff{invmes} we see that the left or right invariant measure differ from the bilateral one only in a normalization constant:
\begin{equation}
\label{eq:rr1}
\frac{\pia x}{\pia y} \;=\; \left\{\begin{array}{rl}
\displaystyle\frac{\pial x}{\pial y} & x\in\LB b0\\[10pt]
\displaystyle\frac{\piar x}{\piar y} & x\in\LB 0a\;.
\end{array}\right.
\end{equation}

The behavior for the full well is proven on the basis of the following comparison inequalities.  
\begin{proposition}\label{prop:comp}
Let $\xa$ be a birth-and-death process and $\xal$, $\xar$ its left and right half-well versions.  Then,
\begin{eqnarray}
\label{eq:rr11}
\EE{\Ta a\ze} = \EE{\Tar a\ze} \quad &,& \quad\var\Bigl(\frac{\Ta{a}{\ze}}{\EE{\Ta a\ze}}\Bigr) = \var\Bigl(\frac{\Tar{a}{\ze}}{\EE{\Tar a\ze}}\Bigr)\;, \\
\label{eq:rr12}
\EE{\Ta b\ze} = \EE{\Tal b\ze} \quad &,&\quad\var\Bigl(\frac{\Ta{b}{\ze}}{\EE{\Ta b\ze}}\Bigr) = \var\Bigl(\frac{\Tal{b}{\ze}}{\EE{\Tal b\ze}}\Bigr)\;, 
\end{eqnarray}
and
\begin{eqnarray}
\label{eq:rr13}
\EE{\Ta \ze{a}} &=& \frac{\pia{\LB b{-1}}\,\EE{\Tar a\ze} + \EE{\Tar \ze{a}}}{\pia{\LB 0a}}\;\ge\; \EE{\Tar \ze a}\;,\\[10 pt]
\label{eq:rr14}
\EE{\Ta \ze{b}} &=& \frac{\pia{\LB 1{a}}\,\EE{\Tal b\ze} + \EE{\Tal \ze{b}}}{\pia{\LB b{0}}}\;\ge\; \EE{\Tal \ze b}\;.
\end{eqnarray}
Furthermore, if 
\begin{equation}\label{eq:rr15.1}
C_a\;\bydef\;\frac{\EE{\Ta a\ze}}{\EE{\Ta\ze{a}}} \vee \frac{\EE{\Ta b\ze}}{\EE{\Ta\ze{b}}}\;\le\;1\;,
\end{equation}
then
\begin{equation}\label{eq:rr15}
\frac{\bigl(1-2\,C_a\bigr)}{2}\,\EE{\Ta\ze{a}}\wedge\EE{\Ta\ze{b}} \;\le\; \EE{\Ta \ze{\{a,b\}}}
\;\le\; \EE{\Ta\ze{a}}\wedge\EE{\Ta\ze{b}}\;.
\end{equation}
\end{proposition}
\proof

The identities in \reff{eq:rr11} and \reff{eq:rr12} are an immediate consequence of \reff{eq:rl70}--\reff{eq:rl73} and \reff{eq:rr1}.

\bigskip\par\noindent
\emph{Proof of \reff{eq:rr13} and \reff{eq:rr14}.}
From \reff{eq:rl71}, \reff{eq:rl72} and \reff{eq:rr11},
\begin{eqnarray}\label{eq:rr18}
\EE{\Ta \ze{a}} &=& \sum_{k=0}^{a-1} \frac{\pi(\LB bk)}{p_k\,\pi(k)}\nonumber\\
&=& \sum_{k=0}^{a-1} \bigl[\pi(\LB b{-1})+\pi(\LB 0k)\bigr]\,\frac{1}{p_k\,\pi(k)}\nonumber\\
&=& \pi(\LB b{-1})\,\sum_{k=0}^{a-1}\frac{1}{p_k\,\pi(k)} + \EE{\Tar \ze{a}}\nonumber\\
&=& \pi(\LB b{-1})\,\bigl(\EE{\Ta a\ze}+ \EE{\Ta \ze{a}}\bigr) + \EE{\Tar \ze{a}}\;.
\end{eqnarray}
Passing $\pi(\LB b{-1})\,\EE{\Ta \ze{a}}$ to the left-hand side we arrive to \reff{eq:rr13}.  The proof of \reff{eq:rr14} is analogous.
\bigskip\par\noindent
\emph{Proof of \reff{eq:rr15}.}
The upper bound is an immediate consequence of the fact that $\Ta \ze{\{a,b\}}\le \Ta\ze{a}\,,\,\Ta\ze{b}$.  For the lower bound we observe that from \reff{espappend+-},
\begin{equation}\label{eq:rr19}
\E{\Ta{\ze}{\{a,b\}}} \;=\; \E{\Ta{\ze}{a}}\,\frac{\E{\Ta{a}{b}}}{\E{\Ta{b}{a}}+\E{\Ta{a}{b}}} - 
\E{\Ta{a}{\ze}}\,\frac{\E{\Ta{b}{a}}}{\E{\Ta{b}{a}}+\E{\Ta{a}{b}}}
\end{equation}
Hence, if $\lambda_a\bydef \E{\Ta{a}{b}}/\bigl[\E{\Ta{b}{a}}+\E{\Ta{a}{b}}\bigr]$, 
\begin{eqnarray}\label{eq:rr19.1}
\EE{\Ta{\ze}{\{a,b\}}}&\ge& \EE{\Ta{\ze}{a}}\,\bigl[\lambda_a - C_a\bigr]\nonumber\\
&\ge& \E{\Ta{\ze}{a}}\wedge \EE{\Ta \ze{b}}\,\bigl[\lambda_a - C_a\bigr]\;.
\end{eqnarray}
Analogously, using \reff{espappend+-1} we obtain
\begin{equation}\label{eq:rr20}
\EE{\Ta{\ze}{\{a,b\}}} \;\ge\; \EE{\Ta{\ze}{a}}\wedge \EE{\Ta \ze{b}}\,\bigl[(1-\lambda_a) - C_a\bigr]\;.
\end{equation}
Adding \reff{eq:rr19.1} to \reff{eq:rr20} we obtain the lower bound in \reff{eq:rr15}.\qed

\subsection{Proof of cut-off and escape behavior for the full well}\label{Proofuffwell}
In this section the notation ``$o(1)$'' stands for a positive term that goes to zero as $a$ goes to infinity.
\\

\bigskip\par\noindent
\emph{Cut-off behavior of $\Ta a\ze$ and $\Ta b\ze$.}
The laws of these hitting times depend only on half-well trajectories, therefore their cut-off behavior has been proven in Section \ref{HWmodel}. 

More formally, let us prove the cut-off behavior of $\Ta a\ze$, the proof for $\Ta b\ze$ is analoguous.\\
We suppose that $\xa$ has a strong drift to the left ($0^+$-SD). We thus have $$K_a^2/\EE{\Ta{a}{\ze}}\linf{a}0.$$

As a consequence of \reff{eq:rr1} the quantities $K_a$ and $\qa$ do not depend on the left part of the well (that is: $\hat K_a=K_a$ and $\hat Q^{(a)}=\qa$). Thus, by \reff{eq:rr11} and since $\qa \;\le\; K_a^2\,K_q$ (see \reff{eq:rl81}), we can apply part (ii) of Lemma \ref{lem:cohw} to the half-well version $\xar$, to conclude that
$\var \bigl(\Ta{a}{\ze}/\EE{\Ta{a}{\ze}}\bigr)\longrightarrow 0$ as $a\to\infty$.  Cut-off follows from Proposition \ref{condsuffcut}.

\bigskip\par\noindent
\emph{Escape behavior of $\Ta\ze a$ and $\Ta\ze b$.}  We consider $\Ta\ze a$, the proof for $\Ta\ze b$  is analogous.  We apply Theorem \ref{thmeta} by proving the validity of conditions 1 and 2 of Definition \ref{condmeta} for $\low=0$ and $\high=\{a\}$.  Condition 1 requires that 
\begin{equation}\label{eq:rr21}
\frac{\EE{\Ta{a}{\ze}}}{\EE{\Ta\ze a}} \vee \frac{\EE{\Ta{b}{\ze}}}{\EE{\Ta\ze a}}\;\linf a0\;.
\end{equation}
But the left identity in \reff{eq:rr11} and the inequality in \reff{eq:rr13} imply that
\begin{equation}\label{eq:rr22}
\frac{\EE{\Ta{a}{\ze}}}{\EE{\Ta\ze a}} \;=\; \frac{\EE{\Tar{a}{\ze}}}{\EE{\Ta\ze a}}
\;\le\; \frac{\EE{\Tar{a}{\ze}}}{\EE{\Tar\ze a}}
\end{equation}
which tends to zero as $a$ diverges because of part (i) of Lemma \ref{lem:cohw} (using again \reff{eq:rr1} and \reff{eq:rl81} as in the previous proof).  On the other hand, 
the leftmost hypothesis in \reff{eq:rlns1} guarantees that $\EE{\Ta{b}{\ze}}/\EE{\Ta\ze a}$ tends to $0$ as $a$ goes to infinity.  This proves \reff{eq:rr21}.  Condition 2 follows from \reff{eq:rlgl}  which implies that
\begin{eqnarray}\label{eq:rr23}
\frac{\EE{(\Ta\ze a)^2}}{\EE{\Ta\ze a}^2} &\le& 2 + 2\,\frac{\EE{\Ta b\ze}}{\EE{\Ta\ze a}}\nonumber\\
&=& 2 + o(1)\;.
\end{eqnarray} 
The second line is, again, due to the leftmost hypothesis in \reff{eq:rlns1}.  This proves that the sequence $ \Ta\ze a/\EE{\Ta\ze a}$ is uniformly square integrable and therefore uniformly integrable.
 
\bigskip\par\noindent

\emph{Escape behavior of $\Ta\ze{\{a,b\}}$.}
We verify conditions 1 and 2 of Definition \ref{condmeta} for $\low=0$ and $\high=\{a,b\}$. For the former we must consider the ratio
\begin{equation}\label{eq:rr24}
R^{(a)} \;\bydef\; \frac{\EE{\Ta a\ze} \vee \EE{\Ta{b}{\ze}}}{\EE{\Ta\ze{\{a,b\}}}}\;.
\end{equation}
By the lower bound in \reff{eq:rr15},
\begin{equation}\label{eq:rr25}
R^{(a)} \;\le\; \frac{2}{1-2C_a} \, \frac{\EE{\Ta a\ze} \vee \EE{\Ta{b}{\ze}}}{\EE{\Ta \ze a} \wedge \EE{\Ta\ze{b}}}\;.
\end{equation} 
As seen in (\ref{eq:rr22}): $\EE{\Ta a\ze}\,/\,\EE{\Ta \ze a} \leq \EE{\Tar a\ze}\,/\,\EE{\Tar \ze a}$ which tends to zero by part (i) of Lemma \ref{lem:cohw}. Analoguously, using the left identity in \reff{eq:rr12} and the inequality in \reff{eq:rr14}, we get that $\EE{\Ta b\ze}\,/\,\EE{\Ta \ze b}$ tends to zero, and as a consequence $C_a$ also tends to zero as $a\to\infty$.\\
Condition 1 is proven, since by hypothesis \reff{eq:rlns1} $\EE{\Ta a\ze}\,/\,\EE{\Ta \ze b}$ and $\EE{\Ta b\ze}\,/\,\EE{\Ta \ze a}$ tend to zero.\\

To prove condition 2, we use the fact that $(\Ta\ze{\{a,b\}})^2\le (\Ta \ze a)^2\,,\,(\Ta\ze b)^2$ and the lower bound in \reff{eq:rr15} to obtain 
\begin{eqnarray}\label{eq:rr27}
\frac{\EE{(\Ta\ze{\{a,b\}})^2}}{\EE{\Ta\ze{\{a,b\}}}^2} &\le& \left(\frac{2}{1-2C_a}\right)^2 \, \frac{\EE{(\Ta \ze{a})^2} \wedge \EE{(\Ta{\ze}{b})^2}}{\EE{\Ta \ze a}^2 \wedge \EE{\Ta\ze{b}}^2}
\nonumber\\
&\le& \left(\frac{2}{1-2C_a}\right)^2  \, \frac{\EE{(\Ta \ze{a})^2}}{\EE{\Ta \ze a}^2} \vee \frac{\EE{(\Ta \ze{b})^2}}{\EE{\Ta \ze b}^2} \\
&\le& \left(\frac{2}{1-o(1)}\right)^2 \, \bigl[2 + o(1)\bigr]\;,\nonumber
\end{eqnarray}
where the last line is due to \reff{eq:rr23} and its analogous for $a\leftrightarrow b$.  This proves the uniform square integrability ---and, in consequence, the uniform integrability--- of the sequence $\Ta\ze{\{a,b\}}/\EE{\Ta\ze{\{a,b\}}}$. 
\qed

\begin{remark}\label{rem:rr1}
The same argument used in \reff{eq:rr22} show that the conditions
\begin{equation}
\label{eq:rr28}
 \frac{\EE{\Tal{b}{\ze}}}{\EE{\Tar{\ze}a}} \;\linf a0\quad ,\quad  \frac{\EE{\Tar{a}{\ze}}}{\EE{\Tal{\ze}b}} \;\linf a0\;.
\end{equation}
are sufficient for the validity of \reff{eq:rlns1}.  This requirement is weaker than \reff{eq:rlns}.
\end{remark}
\section{Proof of part 1 of Theorem \ref{theo:0}: Equality of the laws of direct and reverse trajectories:}\label{ProofRevTraj}

This property is a direct consequence of the reversibility of the invariant measure, and is unrelated to the particular behavior of hitting times.  While we do not doubt that the result is well known, we have not found it spelled out in the literature.  For completeness we are presenting a detailed proof.  The property stated in Theorem \ref{theo:0} is a particular instance of the following proposition.

\begin{proposition}\label{pro:rr2}
Let us consider a recurrent Markov process on a finite alphabet $\alff$ defined by transition probabilities $p(u,v)$ with invariant reversible measure $\pi$. Then, for any $x,y\in\alff$ with $\pi(x),\pi(y)\neq 0$ any natural $k$,
\begin{equation}\label{eq:rs71}
\pr{\TT =k} \;=\; \pr{\TTr =k}
\end{equation}
where $\TT$ is the time needed to hit $y$ after hitting $x$ (defined in Section \ref{ssect:rr1} ). 
\end{proposition}

\proof
It is decomposed in two claims, the first of which is adapted from~\cite{Schpatesc}.   Let 
$v^k=(v_1,\ldots,v_k)$ denote a trajectory of length $k$, let $\vvv$ denote the set of all trajectories of length $k$ ``strictly between $x$ and $y$'' and let $\uuu$ be the set of all finite such trajectories: 
\begin{equation}
\vvv \bydef \Bigl\{v^k : v_1=x; v_k=y;  v_i\neq x, y,  1<i<k\Bigr\}\quad,\quad \uuu \bydef \bigcup_{k<\infty} \vvv \;.
\end{equation}
Let $X_z$ denote the Markov process started at $z\in\alff$.  Following~\cite{Schpatesc}, we say that $X_z$ \emph{starts as } $v^k$ if for all $i=1,\ldots,k$, $X_z(i-1)=v_i$, and for a set $\mathcal{V}\subset\alff$, that $X_z$ starts in $\mathcal{V}$ if $X_z$ \emph{starts as } $v^k$ for some $v^k \in \mathcal{V}$.
Let $\rev$ denote the time-reversal operator on the space of finite trajectories: $\rev(v^k)=(v_k,\ldots,v_1)$.
\bigskip\par\noindent
\emph{Claim 1:} 
\begin{eqnarray}\label{hamp.1}
\lefteqn{\pr{ X_x \mbox{ starts in } \vvv \Bigm{|} X_x  \mbox{ starts in } \uuu }}\nonumber\\
&&\qquad\qquad=\; \pr{ X_y  \mbox{ starts in } \rev(\vvv) \Bigm{|}X_y \mbox{ starts in } \rev(\uuu) }\,.
\end{eqnarray}
Indeed, by reversibility,
\begin{align}\label{hamp.r1}
\pi(x)\,\pr{X_x \; \mbox{starts as} \; v^k}
	&=\pi(x) \p{x}{v_2}\,\p{v_2}{v_3}\cdots\p{v_{k\mtw}}{v_{k\mo}}\,\p{v_{k\mo}}{y} \nonumber\\
	&=\p{v_2}{x}\,\p{v_3}{v_2}\cdots\p{v_{k\mo}}{v_{k\mtw}}\,\p{y}{v_{k\mo}}\,\pi(y) \nonumber\\
	&=\pi(y)\,\pr{X_y \; \mbox{starts as} \; \mathcal{R}(v^k)} \;.
\end{align}
Summing over $v^k \in \vvv$, we get $\pi(x)\,\pr{X_x \; \mbox{starts in} \; \vvv}=\pi(y)\,\pr{X_y \; \mbox{starts as} \; \mathcal{R}(\vvv)}$ and thus, summing over $k$,
\begin{equation}\label{hamp.r2}
\pi(x)\,\pr{X_x \; \mbox{starts in} \; \uuu}=\pi(y)\,\pr{X_y \; \mbox{starts in} \; \mathcal{R}(\uuu)}\;.
\end{equation}
Dividing each side of \reff{hamp.r1} by the corresponding side of \reff{hamp.r2} we arrive to \reff{hamp.1}.

\bigskip\par\noindent
\emph{Claim 2:}\ 
\begin{eqnarray}\label{Ttilde}
\pr{\TT=k}&=&\pr{ X_x \; \mbox{starts in} \; \vvv \bigm{|} X_x \; \mbox{starts in} \; \uuu }\;,\\
\label{Ttilde1}
\pr{\TTr=k}&=&\pr{X_y \; \mbox{starts in} \; \rev(\vvv) \bigm{|} X_y \; \mbox{starts in} \; \rev(\uuu) }\;.
\end{eqnarray}
Let $\tau_x \bydef \tau^{(x)}_{x\to x,y}$ be the time of last visit to $x$ before hitting $y$, and for $n\in\mathbb{N}$ let us denote $\theta_n:\mathbb{N}\to\mathbb{N}$ the ``forward time-translation'' $\theta_n(t)=t+n$.  Partitioning over all possible values of $\tau_x$ and using the Markov property we obtain
\begin{align}\label{hamp.r3}
\pr{\TT=k} 
		&=\sum_n \prr{\tau_a=n;\,X_x\circ\theta_n \; \mbox{starts in} \; \vvv} \nonumber\\
		&=\sum_n \prr{X_x(t)\neq0, t<n; X_x(n)=a; \,X_x\circ\theta_n \; \mbox{starts in} \; \vvv} \nonumber\\
		&=\prr{ X_x \; \mbox{starts in} \; \vvv}\,\sum_n \prr{X_x(t)\neq0, t<n; X_x(n)=a} \;.
\end{align}
Summing over $k$, we get
\begin{equation}
1\;=\;\sum_n\, \prr{X_x(t)\neq0, t<n; X_x(n)=a}\;\sum_k\,\prr{ X_x \; \mbox{starts in} \; \vvv} \;;
\end{equation}
therefore
\begin{equation}\label{hamp.r4}
\sum_n \prr{X_x(t)\neq0, t<n; X_x(n)=a}\;=\;\frac{1}{\prr{ X_x \; \mbox{starts in} \; \vvv}}\;.
\end{equation}
The combination of \reff{hamp.r3} and \reff{hamp.r4} yields \reff{Ttilde}.  The proof of \reff{Ttilde1} is analogous. \qed


{\bf Acknowledgements} 

It is a pleasure to thank Anton Bovier, Olivier Durieu, Aernout van Enter, Antonio Galves, Nicolas Lanchier, Veronique Gayrard and Yuval Peres for enlightening discussions and helpful criticism.  R.F.~wishes to acknowledge the hospitality of Eurandom, the University of Leiden, the University of Groningen  during the completion of this work. Part of the work of R.F. was done during the authors'stay at Institut Henri Poincar\'e, Centre Emile Borel (whose hospitality is acknowledged), for the semester ``Interacting Particle Systems, Statistical Mechanics and Probability Theory''. J.B. wishes to acknowledge the hospitality of  Laboratoire de Math\'ematiques Rapha\"el Salem UMR 6085 CNRS-Universit\'e de Rouen.  J.B. was partially supported by Fondecyt Project 1060485,  Millennium Nucleus Information and Randomness ICM P04-069-F and Programa Basal, CMM. U. de Chile. O.B. wishes  to thank a French-Brazilian agreement CAPES-COFECUB and the European Science Fundation for travel support.


\bibliographystyle{plain}
\bibliography{Biblio-cut-meta}

\end{document}